\documentclass[]{aa}
\usepackage{natbib}
\bibpunct{(}{)}{;}{a}{}{,}
\usepackage{graphicx}
\usepackage{txfonts}
\newcommand{\kms}{km~s$^{-1}$}
\newcommand{\Msun}{$M_\odot$}
\newcommand{\Teff}{T_{\rm eff}}
\newcommand{\CS}{CS~30322-023}
\begin{document}
\titlerunning{CS 30322-023: an ultra metal-poor TP-AGB star}
  \title{CS 30322-023: an ultra metal-poor TP-AGB star?\thanks{Based 
on observations performed with the Very Large
  Telescope at the European Southern Observatory, Paranal, Chile,
  under programs 69.D-0063(A), 73.D-0193(A,B), 74.D-0228(A),75.D-0120(A),76.D-0451(A),165.N-0276(A)}}
  
   \author{T. Masseron \inst{1}
          \and
          S. Van Eck \inst{2}
          \and
          B. Famaey \inst{2}
 \and S. Goriely \inst{2}
 \and B. Plez   \inst{1} 
 \and L. Siess \inst{2} 
 \and T.~C. Beers \inst{3}
 \and F. Primas \inst{4}
\and  A. Jorissen \inst{2}
          }

  \offprints{S. Van Eck}

   \institute{ GRAAL, UMR 5024 CNRS, Universit\'e de Montpellier-II, France 
\and
Institut d'Astronomie et d'Astrophysique, Universit\'e Libre de
Bruxelles, CP 226, Boulevard du Triomphe, B-1050 Bruxelles, Belgium 
\and
Department of Physics and Astronomy, Center for the Study of Cosmic Evolution (CSCE) 
and Joint Institute for Nuclear 
Astrophysics (JINA), Michigan State University, East Lansing, MI 48824-1116, 
USA
\and
European Southern Observatory, K. Schwarzschild Stra\ss e 2, D-85748 Garching 
bei M\"unchen
}

   \date{Received January 3, 2006 /  Accepted May 5, 2006}

\abstract{The  remarkable properties of \CS\ became apparent 
during the course of a high-resolution spectroscopic study of a sample of 23 carbon-enhanced, metal-poor
(CEMP) stars.}
{This sample is studied in order to gain a better understanding of
s- and r-process nucleosynthesis at low metallicity, and to
investigate the role of duplicity.} 
{High-resolution UVES spectra have been obtained, and abundances are derived
using 1-D, plane-parallel OSMARCS models under the LTE
 hypothesis. The derived atmospheric parameters and observed abundances are compared to
evolutionary tracks and nucleosynthesis predictions to 
infer the evolutionary status of \CS.}
{\CS\ is
remarkable in having the lowest surface gravity  ($\log g \le -0.3$)  among
the metal-poor stars studied to date.  As a result of its rather low
temperature (4100~K), abundances could be derived for 35 chemical
elements; the abundance pattern of \CS\ is one of the
most well-specified of all known extremely metal-poor stars.  With [Fe/H]$ = -3.5$, 
\CS\ is the most metal-poor star to exhibit a clear s-process signature,
and the most metal-poor ``lead star'' known. The available evidence
indicates that \CS\ is presently a thermally-pulsing asymptotic giant branch (TP-AGB)
star, with no strong indication of binarity thus far (although a signal of period
192~d is clearly present in the radial-velocity data, this is likely
due to pulsation of the stellar envelope).
Low-mass TP-AGB stars are not expected to be exceedingly rare in a 
magnitude-limited sample such as the HK survey, because their high luminosities 
make it
possible to sample them over a very large volume. 
The strong N overabundance and the low $^{12}$C/$^{13}$C ratio (4) 
in this star is typical of the operation of the CN cycle. 
Coupled with a Na overabundance and the absence of a strong C overabundance, this
pattern seems to imply that hot-bottom burning operated in this star,
which should then have a mass of at least 2~\Msun. However, the luminosity associated with
this mass would put the
star at a distance of about 50~kpc, in the outskirts of the galactic
halo, where no recent star formation is expected to have taken place.
We explore alternative scenarios in which the observed abundance pattern results
from some mixing mechanism yet to be identified occurring in a single
low-metallicity 0.8~\Msun\ AGB star, or from pollution by matter from an 
intermediate-mass AGB companion which has undergone hot-bottom burning. We stress, 
however, that our abundances may be subject to uncertainties due to NLTE or 
3D granulation effects which were not taken into consideration. 
 }
{}
\keywords{Stars: AGB and post-AGB -- Stars: carbon -- Stars: evolution
  -- Stars: individual: CS 30322-023 -- Galaxy: halo}

\maketitle
%
%________________________________________________________________

\section{Introduction}

The star \object{CS~30322-023} first appeared in the \citet{Norris-99} list of
broadband photometry for candidate metal-poor stars selected from the HK survey
\citep{Beers-85,Beers-1992}. Follow-up medium-resolution spectroscopy of \CS\
revealed that it was likely to be rather C-rich. It was therefore included in
our sample of 23 carbon-enhanced metal-poor (CEMP) stars observed with the UVES
spectrograph on VLT-Kueyen, in order to look for more lead stars after the
initial discoveries by \citet{VanEck-01,VanEck-03}. Lead stars are
low-metallicity objects where the operation of the s-process nucleosynthesis was
so efficient that it converted a relatively large fraction of iron-seed nuclei
all the way to lead \citep{Goriely-00}. The sample we investigated was collected
from three different sources:

\begin{itemize}
\item CEMP objects from the HK survey
   not included in
  high-resolution follow-up studies by other teams;
\item Stars from the list of faint high-latitude carbon stars
  in the Hamburg-ESO (HE) survey \citep{Christlieb-01} which have strong
   CN or C$_2$ bands;
\item The catalogue of low-metallicity stars with strong CH bands from    
\citet{Sleivyte-90}.
\end{itemize} 

Preliminary analyses of this sample have been presented by 
\citet{Masseron-04,Masseron-2006}; a more detailed paper is in
preparation as part of Masseron's Ph.D. thesis ({\it Observatoire de Paris},
France).   

The star \CS\ deserves, however, a dedicated analysis, as it exhibits several
pecularities as compared to other metal-poor stars. First, as will be shown in
Sect.~\ref{Sect:abundances}, it is another example of a lead star. Secondly,
despite the presence of a definite, albeit moderate, s-process signature, it has
almost no C enhancement, the strength of the CN bands being due to a strong N
overabundance. It is therefore one of the few known Nitrogen-Enhanced Metal-Poor
(NEMP) stars. Thirdly, its most intriguing property is, by far, its very low
gravity ($\log g$ about $-0.3$), coupled with a rather low temperature (4100~K;
Sect.~\ref{Sect:parameters}). Its evolutionary status is thus puzzling
(Sect.~\ref{Sect:evol}); is it a low-mass ($\sim 0.8$~\Msun) star close to the
tip of the AGB or a rather luminous intermediate-mass star located in the
outskirts of the galactic halo? Either possibility raises intriguing questions
concerning its nature, which are addressed in Sects.~\ref{Sect:proba} and
\ref{Sect:discussion}. Among these are: What is the probability of catching a
low-mass star in an evolutionary stage so close to the tip of the AGB? Is the
star formation history in the halo consistent with late formation of an
extremely low-metallicity intermediate-mass star in the outskirsts of the
Galaxy? Is it a star accreted during a merger event with a satellite galaxy? Is
it a star that evaporated from a globular cluster? Is it a low-mass star
polluted by an intermediate-mass companion formerly on the thermally-pulsing
aymptotic giant branch?  

\section{Observations and radial-velocity variations}
\label{Sect:observations}

Several high-resolution spectra of CS~30322-023 have been obtained at the
European Southern Observatory with the UVES spectrograph mounted on the
VLT-Kueyen telescope, and with the FEROS spectrograph mounted on the 2.2m
telescope (courtesy S. Lucatello). The observation log is listed in
Table~\ref{Tab:Vr}. The first and last spectra have been obtained in the
framework of other UVES programmes \citep[such as the ESO Large Programme on
``First Stars''; e.g.,][]{Spite-2006}, and the corresponding radial velocities
have been kindly communicated to us by M. Spite and J. Johnson. The two spectra
exposed for 3600~s on the blue arm of UVES (two times 1800~s on the red arm)
were used to derive the abundances (Sect.~\ref{Sect:abundances}). Their
signal-to-noise (SNR) ratio is around 200/1 at 400~nm. The spectra exposed for
90~s were only used to derive radial velocities by cross-correlating the
observed spectrum with the two high-SNR spectra. The adopted heliocentric
velocity for a given low-SNR spectrum is the average of these two differential
values; it is listed in Table~\ref{Tab:Vr}. Arcturus and solar templates were
used to estimate the internal consistency of the radial-velocity measurement,
and a telluric template spectrum was used to evaluate the zero point offset
velocity. The uncertainty on the radial velocity is the root mean square of the
template error and the precision of the centering of the gaussian fit to the
cross-correlation profile. Both uncertainty sources, along with their root mean
square error, are listed in Table~\ref{Tab:Vr}.

The standard deviation of the observed radial velocities amounts to
1.24~\kms, which is considerably larger than the average uncertainty, amounting
to only 0.18~\kms. Possible causes for these variations are investigated in 
Sect.~\ref{Sect:Vr}.

\begin{table*}
\caption{Observation log. The uncertainty on the radial-velocity $Vr$ arising from the template 
mismatch ($\epsilon_{temp}$), from the gaussian fit of the correlation profile
($\epsilon_{fit}$), and their 
root-mean-square ($\epsilon_{tot}$), are also listed. 
\label{Tab:Vr}
}
\begin{tabular}{llllllllll}
\hline 
\hline Date & HJD &  Exp. time & $Vr$ & $\epsilon_{temp}$ & $\epsilon_{fit}$ & $\epsilon_{tot}$ & Telescope &
Programme \\ 
(YYYY/MM/DD) & (-2\ts450\ts000) & (s)  & (km/s) & (km/s) & (km/s) & (km/s)\\  
\hline
2001/11/04 & 2218.49762 & 3100     & 116.53$^a$ & -    & 0.08 & 0.08 & VLT/UVES & 165.N-0276(A)\\
2002/04/24 & 2388.83689 & 3600     & 116.40     & 0.04 & 0.04 & 0.06 & VLT/UVES & 069.D-0063(A)\\ 
2002/05/11 & 2405.86677 & 3600     & 114.59     & 0.05 & 0.04 & 0.06 & VLT/UVES & 069.D-0063(A)\\  
2004/04/29 & 3124.92121 & 90       & 115.92     & 0.04 & 0.04 & 0.06 & VLT/UVES & 073.D-0193(B)\\ 
2004/06/14 & 3170.81670 & 90       & 115.75     & 0.04 & 0.03 & 0.05 & VLT/UVES & 073.D-0193(B)\\
2004/08/17 & 3234.64390 & 90       & 112.81     & 0.05 & 0.02 & 0.05 & VLT/UVES & 074.D-0228(A)\\ 
2004/11/11 & 3320.53675 & 90       & 115.71     & 0.05 & 0.03 & 0.06 & VLT/UVES & 074.D-0228(A)\\ 
2005/06/29 & 3550.89234 & 900      & 115.09$^b$ & 0.37 & 0.05 & 0.37 & 2.2m/FEROS & 075.D-0120(A)\\
2005/07/26 & 3577.86729 & 900      & 114.61$^b$ & 0.39 & 0.08 & 0.40 & 2.2m/FEROS & 075.D-0120(A)\\
2005/07/27 & 3578.81266 & 900      & 113.69$^b$ & 0.29 & 0.09 & 0.30 & 2.2m/FEROS & 075.D-0120(A)\\
2005/07/28 & 3579.89378 & 900      & 113.82$^b$ & 0.61 & 0.08 & 0.61 & 2.2m/FEROS & 075.D-0120(A)\\
2005/08/17 & 3599.67995 & 2000     & 113.35$^c$ & 0.03 & 0.05 & 0.06 & VLT/UVES & 076.D-0451(A)\\
\hline
\end{tabular}

$^a$ Data kindly communicated by M. Spite\\
$^b$ Data kindly communicated by S. Lucatello\\
$^c$ Data kindly communicated by J. Johnson\\
\end{table*}

Broadband $UBV$ photometry is provided by \citet{Norris-99}, while near-infrared
$JHK$ photometry is available from the {\it Two Micron All Sky Survey}
\citep{2MASS}, as listed in Table~\ref{Tab:phot}. Given the large distance to CS~30322-023 (Sect.~\ref{Sect:evol}), 
the reddening along its line of sight across the Galaxy was taken from
\citet{Burstein-Heiles-1982} to be $E_{B-V} = 0.06$, from which it follows that
$A_V = 0.18$~mag. This value is close to the 0.16~mag predicted by the model of 
\citet{Arenou-92}.  The reddening law $A_\lambda/A_V$ from \citet{Cohen-1981}, appropriate for
carbon stars, has been applied to deredden the observed magnitudes.
Zero-magnitude fluxes from \citet{Schmidt-Kaler-1982} and
\citet{Cohen-2003} have been used to construct the spectral energy
distribution. The bolometric flux has then been obtained by computing the
integral $\int \lambda F_\lambda \; {\rm d}\ln\lambda$ using the trapezoidal
rule on the $UBV$ and $JHK$ bands. A bolometric correction of $-0.30$~mag is
finally obtained, yielding $m_{\rm bol} = 11.7$. It should be noted that this
bolometric correction differs somewhat from the value of $-0.8$~mag predicted by
Table~6 of \citet{Alonso-1999} for giants with [Fe/H]$ = -3$ and $T_{\rm eff}Ê=
4100$~K.
 
\begin{table}
\caption{Available photometry for CS 30322-023; $m$ and $m_0$ are
the observed and de-reddened magnitudes in the considered 
band, respectively. 
\label{Tab:phot}}
\begin{tabular}{lllll}
\hline
\hline
Band & $m$ & $\pm$ & $m_0$ & Ref. \\
\hline
$U$ & 14.48 & 0.02 & 14.19 & (1) \\
$B$ & 13.49 & 0.03 & 13.24 & (1) \\
$V$ & 12.21 & 0.03 & 12.02 & (1) \\
$J$ & 10.050 & 0.024 & 10.00 & (2) \\
$H$ & 9.479 & 0.024 & 9.45 & (2) \\
$K$ & 9.313 & 0.021 & 9.30 & (2) 
\medskip\\
Bol. &      &       &11.7 & (3)\\
\hline
\end{tabular}

(1) \citet{Norris-99};
(2) \citet{2MASS};
(3) This work
\end{table}

\section{Atmospheric parameters of CS~30322-023}
\label{Sect:parameters}

The effective temperature of CS~30322-023 was obtained, in the usual manner, by
requiring that the abundances derived from various iron lines exhibit no trend
with excitation potential. Gravity was obtained through ionisation balance, and
microturbulence was set by requiring the absence of a trend between abundances
and reduced equivalent widths. A photometric $T_{\rm eff}$ estimate has been
obtained from the calibration of \citet{Alonso-1998}, applied on the 2MASS $J-H$
and $J-K$ indices, assuming a metallicity of $-3.0$. 

\begin{figure}
\includegraphics[width=5cm,height=8cm,angle=-90]{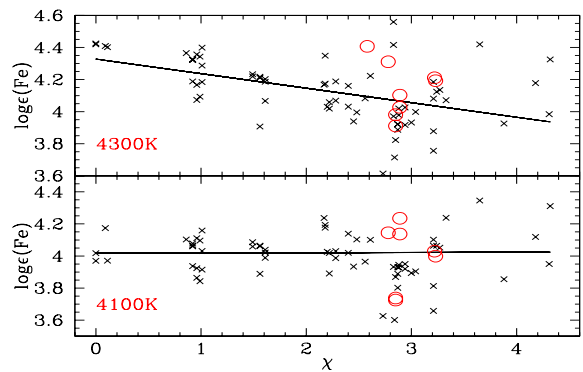}
\caption{\label{Fig:4100_4300}
 Iron abundances derived from all lines measured in \protect\CS\
  (crosses: Fe~I lines; open circles: Fe~II lines),
  as a function of the lower excitation potential 
for the adopted model (Lower panel: $\Teff$
  = 4100~K, $\log g = -0.3$)
and for a 
model of higher temperature and gravity (Upper panel: $\Teff$
  = 4300~K, $\log g = 0.5$).
The latter model -- the best one at 4300~K -- is seen to be less 
satisfactory than the adopted 4100~K model. In both cases,  plane-parallel LTE model 
atmospheres (with [Fe/H] $=-3.5$ and a microturbulent velocity of 2.2~\kms) have been used.
}
\end{figure}

\begin{figure}
\includegraphics[width=5cm,height=8cm,angle=-90]{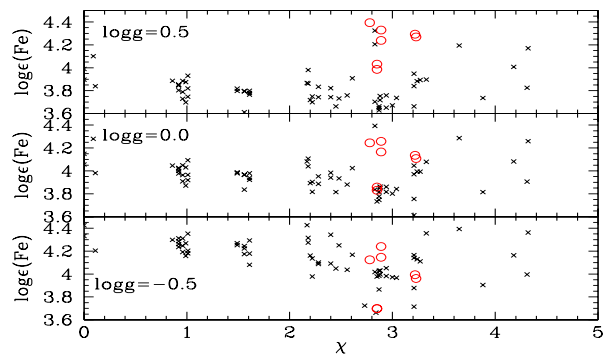}
\caption{\label{Fig:logg}
Same as Fig.~\ref{Fig:4100_4300} for different
  gravities at  $\Teff$
  = 4100~K.       
}
\end{figure}

The derived atmospheric parameters, using model atmospheres belonging to the
OSMARCS family \citep{Gustafsson-03a}, are listed in Table~\ref{Tab:parameters}.
The discrepancy between the photometric and spectroscopic temperature estimates
probably results, on one hand, from the fact that the CS~30322-023 parameters
fall at the edge of the validity domain of Alonso's calibration and, on the
other hand, from model errors impacting the spectroscopic determination, as we
now discuss. Fig.~\ref{Fig:4100_4300} reveals that a temperature $\Teff =
4300$~K, as suggested by the photometric calibration, combined with a higher
gravity, does not yield a satisfactory solution. Fig.~\ref{Fig:logg}, which
presents the Fe~I and Fe~II abundances derived from various lines for
plane-parallel, LTE model atmospheres, shows that the best solution is obtained
for $\Teff = 4100$~K and $\log g = -0.3$. The uncertainty on the gravity, in the
framework of Fe~I/Fe~II ionisation equilibrium under LTE (see, however, the
dicussion about NLTE effects below), may be estimated by comparing the three
panels of Fig.~\ref{Fig:logg}, corresponding to $\log g = 0.5, 0.0$ and $-0.5$.
The gravity of \CS\ clearly lies between $-0.5$ and 0.0. The value $\log g =
-0.3\pm0.3$ has therefore been adopted.

\begin{table*}
\caption[]{\label{Tab:parameters}
Atmospheric parameters of CS~30322-023. In the third column, `PP'
stands for plane-parallel and `S' for spherical (see text for a 
discussion of the uncertainties).
}
\begin{tabular}{lll}
\hline
\hline
$T_{\rm eff}$ (K) & 4100 & excitation balance (PP-model)\\
                  & 4100 & excitation balance (S-model)\\
                  & 4350 & 2MASS photometry\\
$\log g$ & $-0.3\pm0.3$ & Fe~I/Fe~II ionisation balance in PP-model (LTE)\\
         & $< -0.3$ & Fe~I/Fe~II ionisation balance in S-model (LTE)\\
microturbulent
vel. (km/s) & 2.2 & PP-model\\
                 & 1.7 & S-model\\
\hline
\end{tabular}
\end{table*}

We have also attempted to employ spherical atmosphere models, which are better
suited to low-gravity stars than the conventionally used plane-parallel models.
The Fe~I/Fe~II ionisation balance would require spherical models with an
effective gravity $\log g < -0.3$. We could not, however, converge models with
such low surface gravities, due to the combination of the large atmospheric
extension and the scattering character of the radiation field at such low
metallicities, which is decoupled from the local thermal pool of the gas. 

It should be noted that the gravities of metal-poor giants derived from LTE
iron-line analyses are most probably underestimated, because of the neglect of
non-LTE effects \citep{Thevenin-99,Israelian-2001,Israelian-2004,Korn-2003}. We
plan to investigate these effects in a forthcoming paper with models
specifically tailored to \CS. Nevertheless, the conclusion that \CS\ is the most
evolved star found thus far among CEMP stars remains valid, as the comparison
presented in Fig.~\ref{Fig:HRall} involves stars analyzed under LTE as well. In
absolute terms, with NLTE corrections on the gravity of 0.5~dex \citep[as
derived by ][ for giant stars with [Fe/H{]}$ = -3$]{Thevenin-99,Korn-2003} or
even up to 1.0~dex \citep[from the comparison of LTE and NLTE gravities for
CS~29498-043 derived by][]{Aoki-2002,Israelian-2004}, \CS\ would remain on the
TP-AGB (Figs.~\ref{Fig:HRall} and \ref{Fig:HRLionel}).  

\section{Evolutionary status of CS~30322-023}
\label{Sect:evol}

\subsection{Position in the Hertzsprung-Russell diagram}

The gravity and effective temperatures derived in
Sect.~\ref{Tab:parameters} may be combined 
with the apparent bolometric magnitude to derive the distance of
CS~30322-023 as a function of the adopted mass, using the relationship:
\begin{eqnarray}
\log M/M_{\odot} &=& \log g/g_\odot + \log L/L_\odot - 4 \log T_{\rm
 eff}/T_{\rm eff,\odot} \\ 
&=& -2.25 - 0.4 M_{\rm bol}\\ 
&=& 2 \;\log d\;({\rm pc}) - 8.93,
\end{eqnarray}
where we have adopted $T_{\rm eff} = 4100$~K, $\log g = -0.3$ and $m_{\rm bol} =
11.7$ for CS~30322-023 [denoted set (1) in Table~\ref{Tab:masses}]. The
corresponding distances and luminosities are listed in Table~\ref{Tab:masses}.
To illustrate the effects of the uncertainties, values for $\log L/L_{\odot}$,
$M_{\rm bol}$ and $d$ [listed under set (2)] have been derived assuming that the
gravity could be 1~dex higher (due to NLTE effects) and the temperature 250~K
higher. 

\begin{table*}
\caption{Distance of CS~30322-023 as a function of the adopted mass,
  assuming $\log g = -0.3$, $T_{\rm eff} = 4100$~K  and $m_{\rm bol} =
  11.7$ (corresponding to 
$BC_V = -0.3$ and $A_V = 0.18$~mag). Solutions corresponding to this
  set of parameters are listed under the header `set (1)', whereas an
  alternative solution, accounting for uncertainties in gravity
  and temperature, is listed under the header `set (2)' 
($\log g = 0.7$, $T_{\rm eff} = 4350$~K  and $m_{\rm bol} =
  11.7$). 
The ages up to
  the TP-AGB phase are read from Fig.~\ref{Fig:ageLionel} for the 0.8,
  4, and 9~\Msun\ models.  
\label{Tab:masses}}
\begin{tabular}{rllllllllrlll}
\hline
\hline
$M/M_\odot$ && \multicolumn{2}{c}{$\log L/L_\odot$} &&
\multicolumn{2}{c}{$M_{\rm bol}$} && \multicolumn{2}{c}{$d$ (kpc)} &
age (y)\\
\cline{3-4}\cline{6-7} \cline{9-10}
  & set & (1) & (2) && (1) & (2) && (1) & (2) &\\
\hline
10 && 5.15 & 4.24 && $-$8.1 & $-$5.9 && 92 & 33 & - \\
9  && 5.10 & 4.20 && $-$8.0 & $-$5.8 && 87 & 31 & $2.5\;10^7$\\
5 && 4.84 &  3.94 && $-$7.4 & $-$5.1 && 65 & 23 & - \\
4 && 4.75 &  3.85 && $-$7.1 & $-$4.9 && 58 & 21 & $1.6\;10^8$\\
3 && 4.62 &  3.72 && $-$6.8 & $-$4.6 && 50 & 18 & - \\
1 && 4.15 &  3.24 && $-$5.6 & $-$3.4 && 29 & 10 & - \\
0.8&&4.05 &  3.14 && $-$5.4 & $-$3.1 && 26 & 9  & $13.8\;10^9$\\
\hline
\end{tabular}
\end{table*}

\begin{figure}
  \centering \includegraphics[width=8cm]{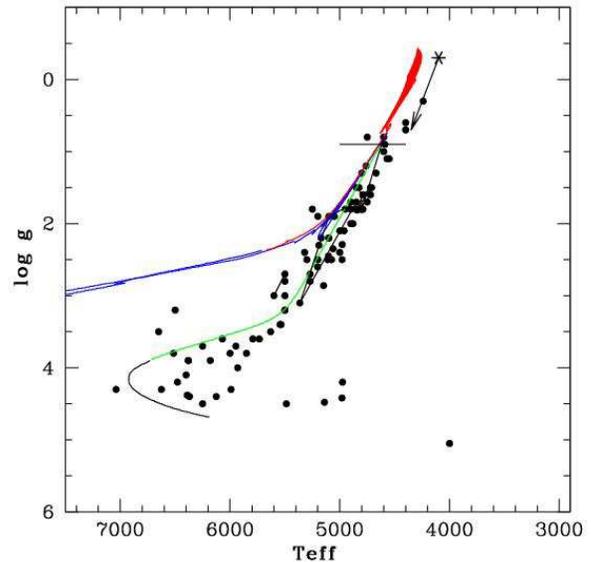}
    \caption{Hertzsprung-Russell diagram for CEMP stars
    collected from the literature (see references in the text), 
    revealing that \protect\CS\ (asterisk) is the most extreme case known. 
The solid line corresponds to the evolutionary track of
a 0.8~\Msun\ model with [Fe/H] $= -3.8$ (see text). The horizontal bar on the
evolutionary track marks the tip of the RGB. The arrow corresponds 
to a conservative 250~K uncertainty on $\Teff$ and a 1.0~dex uncertainty on the gravity of 
\protect\CS\, due to the use of
    LTE instead of NLTE model atmospheres (see text).  
}
  \label{Fig:HRall}
\end{figure}

\CS\ appears to be the star with the lowest gravity detected so far
among CEMP stars, as shown in Fig.~\ref{Fig:HRall}, which collects all such stars
from the literature \citep{Dearborn-86,Ryan-1991,McWilliam-95b,Norris-97,
Norris97b,Sneden-98,Gratton-2000,Hill-2000,Sneden-00,Westin-00,Norris-2001,
Preston-Sneden-01,VanEck-01,Travaglio-2001,Aoki-2001,Christlieb-2002,Hill-02,
Johnson-02,Aoki-2002c,Aoki-2002b,Aoki-2002a,Aoki-2002,Aoki-2003,Cohen-2003a,
Lucatello-2003,Sneden-03a,VanEck-03,Aoki-2004,Cohen-2004,Honda-2004,
Johnson-Bolte-2004,Simmerer-2004,Barbuy-2005,Plez-Cohen-2005,Frebel-2005,
Frebel-2006,Masseron-2006}. 

With $T_{\rm eff} = 4240$~K and $\log g = 0.3$, the star HD~110184 (which,
despite the reported detection of Th, is not considered an r-process enhanced
star, according to the criteria of Beers \& Christlieb 2005; its [Eu/Fe] ratio
is too low) studied by \citet{Honda-2004} is the second most evolved star after
\CS. Then comes \object{CS~29498-043}, studied by \citet{Aoki-2002,Aoki-2004}, 
\citet{Weiss-2004}, and \citet{Israelian-2004}, with $T_{\rm eff} = 4400$~K and  $\log g =
0.6$ \citep[][a subsequent analysis by Aoki et al., 2004, yielded instead
$T_{\rm eff} = 4600$~K and $\log g = 1.2$]{Aoki-2002}. Using a NLTE analysis on
Fe lines, \citet{Israelian-2004} derive a higher gravity ($\log g = 1.5$), thus
indicating that LTE analyses underestimate the gravity by about 1~dex for this
star.

Stars that may also be located above the tip of the red-giant branch (predicted
to occur at $\log g = 0.9$ in a 0.8~\Msun, [Fe/H]$ = -3.8$ model, as indicated
by the horizontal mark on the evolutionary track of Fig.~\ref{Fig:HRall})
also include \object{CS~30314-067}
\citep{Aoki-2002a}, \object{CS~22948-027}  \citep[][but here again Barbuy et
al., 2005, obtain a larger $\log g$ value of 1.8]{Preston-Sneden-01}, and
\object{CS~30301-015} \citep{Aoki-2002}.

\citet{Mauron-2004,Mauron-2005} list about 120 cool carbon stars in the halo
(with $JHK$ colors very similar to those of \CS), located at quite large
galactocentric distances (up to 120~kpc!). They were not plotted in
Fig.~\ref{Fig:HRall}, however, as neither their metallicities nor their
gravities are known.  

\begin{figure}
  \centering \includegraphics[width=8cm]{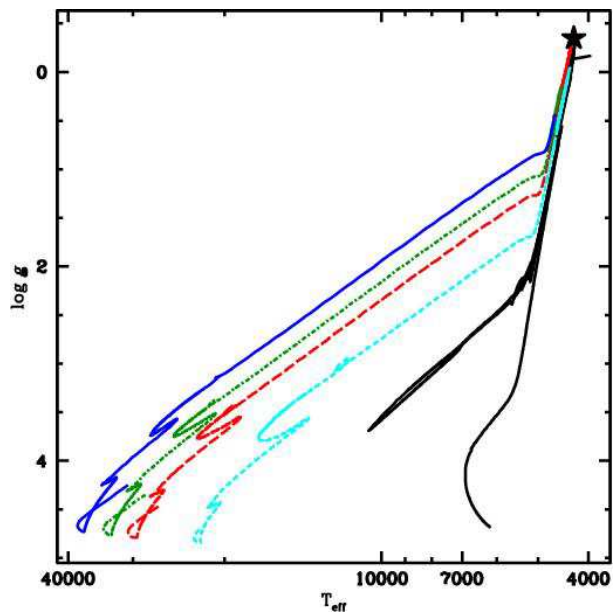}
    \caption{Evolutionary tracks (in a $\log g$ -- $T_{\rm eff}$ plane) 
    of a 0.8~$M_\odot$ model at [Fe/H] $=-3.8$ (black solid
     line) and of a 4~$M_\odot$ (cyan short-dashed line), 7~$M_\odot$
    (red long-dashed
     line), 9~$M_\odot$ (green dot-dashed), and 12~$M_\odot$ (blue
     solid line), at metallicity $Z=10^{-5}$ ([Fe/H]$\simeq -3.3$),
     computed with \scshape STAREVOL\upshape\ \citep{Siess-2006} from the main
     sequence to the thermally-pulsing AGB phase (for the 0.8, 4, 7 and 9
     $M_\odot$ models), and to neon ignition for the 12~$M_\odot$ model.
     The position of \CS\ is represented by the asterisk.}
  \label{Fig:HRLionel}
\end{figure}

The low gravity derived for \CS\ appears, at first sight, quite puzzling. A
comparison with low-metallicity stellar evolution tracks reveals, however, that
both the gravity and effective temperature of CS~30322-023 are in fact typical
of low-metallicity red-giant stars. Fig.~\ref{Fig:HRLionel} displays the
evolutionary tracks in a gravity -- effective temperature plane for a
0.8~$M_\odot$ model of metallicity [Fe/H]$=-3.8$, and for intermediate-mass
stars of 4, 7, 9 and 12~$M_\odot$ and $Z = 10^{-5}$ ([Fe/H]$=-3.3$), computed
with \scshape STAREVOL\upshape\ \citep{Siess-2006} all the way from the pre-main
sequence to the thermally-pulsing AGB stage (for the 0.8, 4, 7 and 9~$M_\odot$
models), and to the ignition of neon for the 12~$M_\odot$ model. For these
computations, we use the \citet{Reimers-1975} mass-loss rate up to the end of
core He burning, and then switch to the \citet{Vassiliadis-Wood-1993}
prescription during the AGB phase. The region around $\log g = 0$ and $\log
\Teff = 3.61$ is reached by all five models, thus providing a non-unique
solution! However, each of these possible solutions presents its own
difficulties. In the case of the 0.8~$M_\odot$ model, the evolutionary
timescales may be a problem. Although the model star has a total lifetime of
13.8~Gyr (about the age of the Galaxy), the time spent by such stars in the
TP-AGB phase might seem prohibitively short to ensure that some are actually
observed. We will show in Sect.~\ref{Sect:proba} that this is not the case,
since the short time span on the TP-AGB is more than offset by the large volume
within which these very luminous stars may be detected. In the case of the
intermediate or massive stars, one may wonder whether such massive short-lived
stars could still be found in the galactic halo, especially since the distance
estimates for \CS\ presented in Table~\ref{Tab:masses} places it at the
outskirts of the halo.

\begin{figure}
   \centering
   \includegraphics[width=8cm]{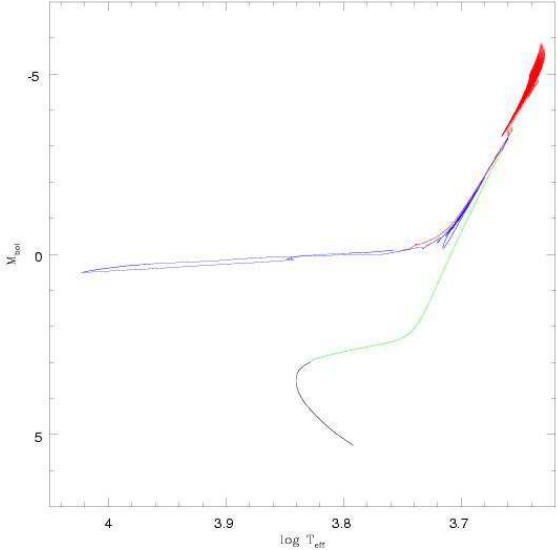}
      \caption{
Same as Fig.~\protect\ref{Fig:HRLionel} for the 0.8~\Msun\ model 
in the usual $M_{\rm bol}$ -- $\Teff$ plane.             }
         \label{Fig:HR-0.8}
   \end{figure}

\subsection{Observational evidence}
\label{Sect:Vr}

\begin{figure}
\begin{center} 
\includegraphics[width=5cm,height=8cm,angle=-90]{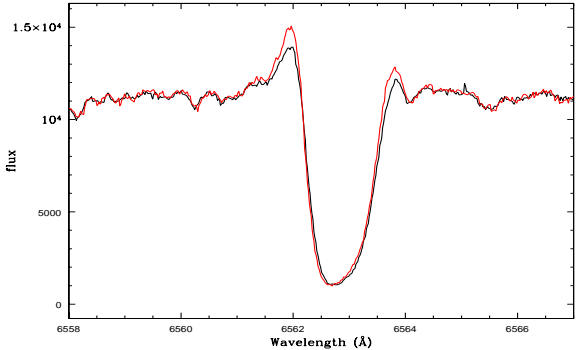}
\caption{Sample spectra around the H$\alpha$ line (the flux is in arbitrary
units), obtained on 2002/04/24 and 2002/05/11. 
The change in the profile over 17 days reflects some kind of activity,
such as, for instance, a strong wind. 
}\label{Fig:Halpha}
\end{center}
\end{figure}

The variability observed for the H$\alpha$ line profile suggests the presence of
a stellar wind (Fig.~\ref{Fig:Halpha}), and provides a further argument in favour
of the very evolved nature of \CS.

Unfortunately, there is no available information on possible optical
variability, which should be present if CS~30322-023 lies high on the AGB. The
profile resulting from the correlation of the spectrum with the Arcturus
template used to derive the radial velocity also does not exhibit the asymmetric
shape expected for Mira variables experiencing shock waves in their photospheres
\citep{Alvarez01}. Nonetheless, as we now show, the available radial-velocity
data (Table~\ref{Tab:Vr}) exhibit properties that appear to be typical of
low-metallicity pulsating stars. 

It was shown in Sect.~\ref{Sect:observations} that the radial velocity standard
deviation (1.24~\kms) is considerably larger than the average uncertainty,
amounting to 0.18~\kms. These variations may be caused either by the pulsation
of the stellar envelope or by an orbital motion in the presence of a companion
(or a combination of both). Let us first stress that the stellar properties
listed in Table~\ref{Tab:masses} put stringent constraints on the admissible
orbital periods. For an $0.8$~\Msun\ star, the stellar radius corresponds to
210~$R_\odot$ with atmospheric parameter set (1) [i.e., $\Teff = 4100$~K and
$\log g = -0.3$], and to 65~$R_\odot$ for set (2) [i.e., $\Teff = 4350$~K and
$\log g = 0.7$]. In a binary system involving two stars of 0.8~\Msun\ (the
primary being the observed low-metallicity star assumed to be a long-lived
low-mass star and the secondary a white dwarf), the minimum admissible orbital
periods corresponding to \CS\ not filling its Roche lobe amount to 1180~d and
200~d, respectively. These periods represent absolute lower limits, since
higher-mass stars have larger radii, and hence longer threshold orbital periods.
 
\begin{figure}
\centering \includegraphics[width=8cm]{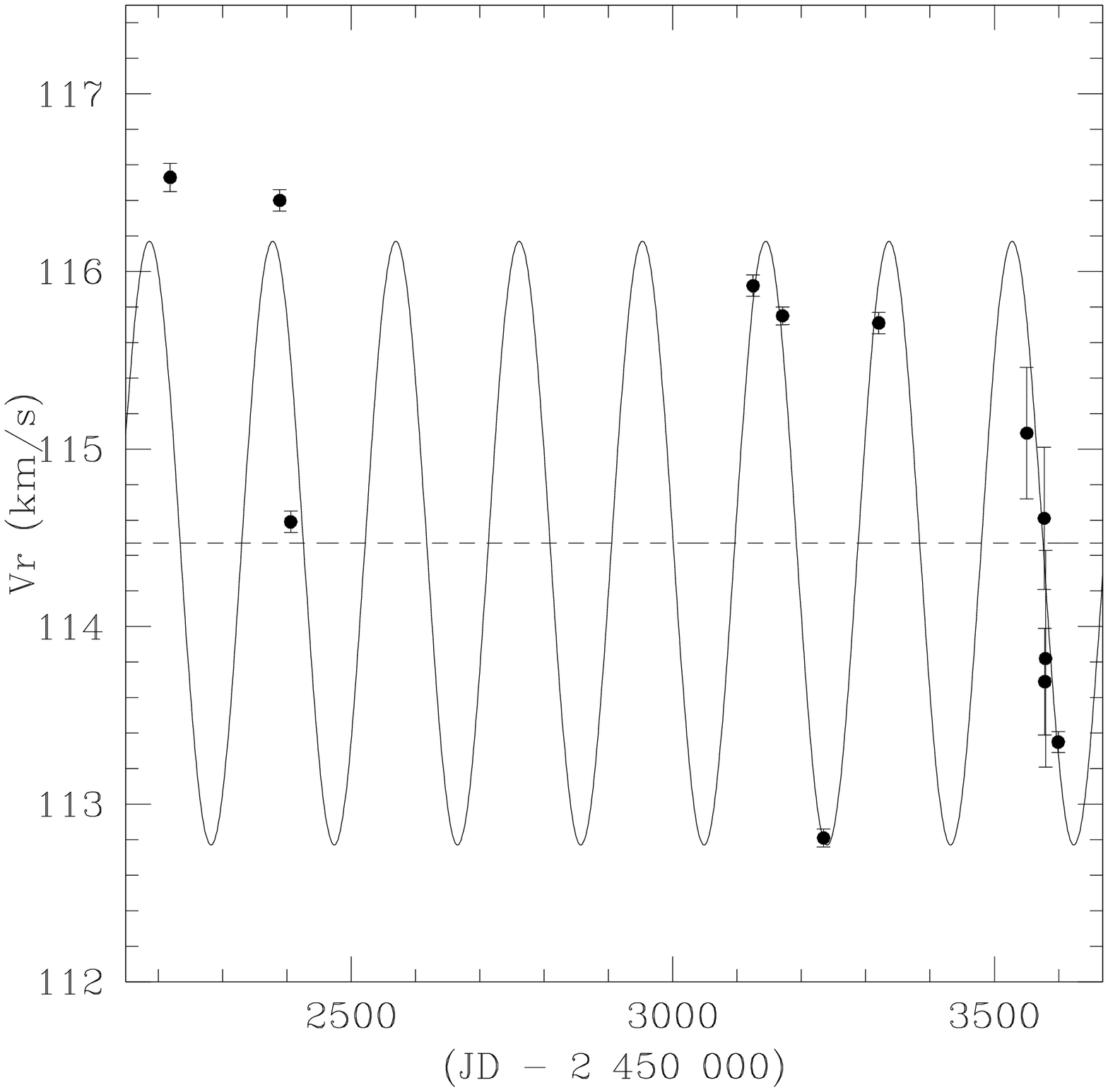}
\centering \includegraphics[width=8cm]{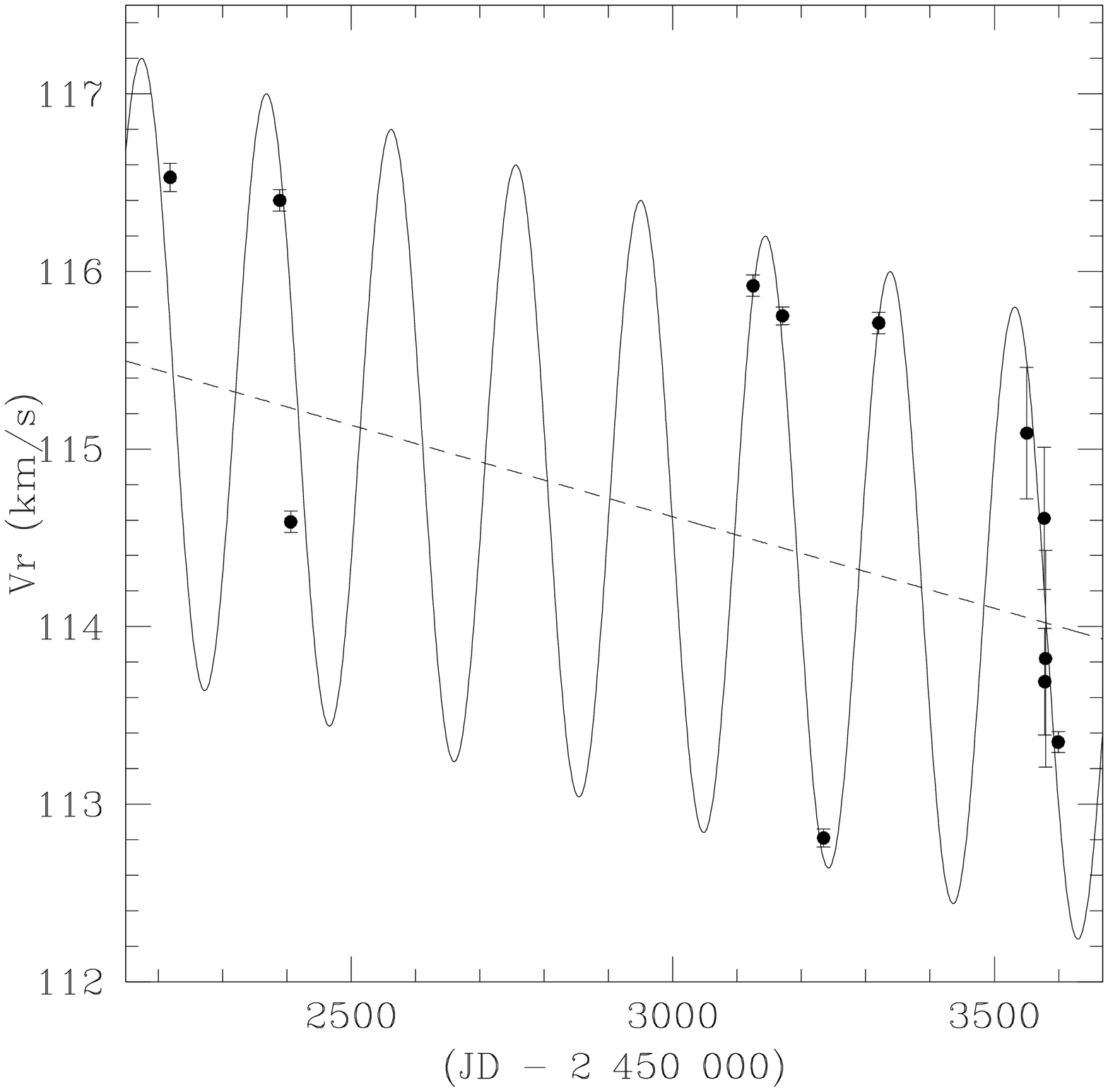}
\caption{\label{Fig:Vr}
Radial velocity as a function of Julian date. 
Top panel: A sine-wave of period 191.7~d, fitted on the data points with
$JD > 2\ts453\ts000$. Lower panel: The same sine-wave superimposed on a
linear trend of 1.0~\kms\ per 1000~d, possibly representing an orbital 
motion.  
}
\end{figure}

Interestingly, a sine-wave of period 192~d very nicely fits the radial-velocity
data at $JD > 2\ts453\ts000$ (upper panel of Fig.~\ref{Fig:Vr}). From the above
considerations, this signal has too short a period to represent an orbital
motion; moreover, it does not fit the earlier data points. It is likely,
therefore, that it represents pulsation of the stellar envelope. The pulsation
periods of low-metallicity stars are shorter than those of solar-metallicity
Miras, and their range matches the 192~d period observed for \CS\
\citep{Alvarez-1997}.  The poor fit to the earlier data points may either
be caused by a sudden phase shift of the (possibly semi-regular) pulsation, or
by a long-term trend associated with a binary motion. 

We now consider the latter possibility. The dashed line in the lower panel of
Fig.~\ref{Fig:Vr} corresponds to a shift of 1.0~\kms\ over 1000~d. It provides a
better fit to the earlier data points; such a trend is not unexpected for
binary systems involving chemically-peculiar red giants such as \CS. To fix the
ideas, let us assume that \CS\ is a binary with a period of 5200~d, as is
typical for barium stars \citep{Jorissen-03}. In this case, our measurements
have only sampled a quarter of the orbital cycle. In the most favourable case,
the observed velocity range is then equal to the velocity semi-amplitude, $K$,
which would amount to 7.2~\kms\ for a circular orbit seen edge-on and involving
two stars of 0.8~\Msun\ (see above). The possible long-term trend of 1.3~\kms\
over 1300~d is not incompatible with the above estimate of $K$, provided that
the orbital plane is inclined on the sky. 

To conclude this discussion, there is no firm support in favour of the binary
nature of \CS, although in this case the absence of evidence is not evidence of
absence, especially since any orbital motion will be very difficult to
disentangle from the large intrinsic jitter. The binary hypothesis will
therefore be left open in the discussion of Sect.~\ref{Sect:discussion}.

%% Radial velocity 
%% variations on time scales on 
%% the order of 100 -- 200~d are nevertheless present, but 
%% these are very likely 
%% related to the intrinsic jitter of the envelope, which is a common phenomenon among evolved 
%% stars, as \CS\ is supposed to be
%% (Sect.~\ref{Sect:evol}). For instance, 
%% intrinsic S stars on the asymptotic giant branch
%% have radial-velocity standard deviations of typically 1.5~\kms, 
%% up to  4~\kms\ in the
%% most extreme cases \citep[Fig.~10 of ][]{VanEck-Jorissen-99c}. \citet{Jorissen-VE-98} 
%% showed (their Fig.~1) that the intrinsic radial-velocity jitter of red giants correlates 
%% with the width of the cross-correlation dip (cc-dip, corrected for the instrumental width), which is 
%% loosely related to luminosity.
%% Since the $\sigma$ of the 
%% gaussian fitted to the UVES cc-dip of \CS\ amounts to 7.4~\kms, the   
%% reduced width of the cc-dip amounts to 6.7~\kms, considering an 
%% instrumental width of 3~\kms\ ({\it i.e.,} a resolution of $100\,000$). For that value of the reduced cc-dip, red giants 
%% appear to have a radial-velocity jitter in the range 1 -- 3~\kms, encompassing the observed 
%% standard deviation of 1.2~\kms\ for \CS. 

%% Thus allowing for this intrinsic jitter, it is nevertheless possible that the radial 
%% velocity data points plotted on Fig.~\ref{Fig:Vr} exhibit a downward trend of about 1.5~\kms\ 
%% over the 1300~d time span of the observations. 

%% Such a trend would certainly be compatible with the expected properties of a binary system 
%% involving \CS. 

\section{Probability of observing an extremely metal-poor TP-AGB star}
\label{Sect:proba}

Since it is very unlikely that a massive or intermediate-mass star has been
formed in a recent star-forming episode at such large galactocentric distances
({$>20$~kpc, Table~\ref{Tab:masses}) in the halo of the Galaxy, in this section
we evaluate the likelihood of an alternative scenario, that CS~30322-023 is a
TP-AGB star formed shortly after the birth of the Galaxy. 

We now estimate the relative fraction of halo TP-AGB stars in a
magnitude-limited sample (such as the HK survey), as compared to main-sequence
turnoff or horizontal-branch stars {\it from a co-eval sample}. The assumption
of co-evality is probably valid as long as one restricts the sample to extremely
metal-poor stars with [Fe/H] $< -3.0$. We further assume that a star of $0.8
M_\odot$ with [Fe/H]$=-3.8$ has a lifetime (up to the planetary nebula phase)
precisely equal to the age of the galactic halo (Sect.~\ref{Sect:evol})
.\footnote{We assume in this section that the lifetimes derived from the stellar
models with [Fe/H]$=-3.8$ apply to \protect\CS, despite the small difference in
metallicity.} 

In an ideal co-eval sample ({\it i.e.}, not subject to observational selection
biases) consisting of the oldest stars in the Galaxy with
$M \ge M_{\rm low}$, $M_{\rm low}$ being the low-mass cutoff of the
population (to be fixed below), the fraction, $F$, of TP-AGB
stars expected in such a sample is:
\begin{equation}
\label{Eq:fractionAGB}
F = \frac{\int_{M_{AGB}}^{0.8 M_\odot} \Psi_{\rm IMF} (M) \; {\rm d} M} 
   {\int_{M_{\rm low}}^{0.8 M_\odot} \Psi_{\rm IMF} (M) \; {\rm d} M},
\end{equation}
where $\Psi_{\rm IMF}$ is the initial mass function. Here we adopt 
$\Psi_{\rm IMF} (M) \propto M^{-1.3}$ for the mass range [0.1,
  0.5]~$M_\odot$, 
and $\Psi_{\rm IMF} (M) \propto M^{-2.3}$ for the mass range 
]0.5, 0.8]~$M_\odot$ \citep{Kroupa-2001}. 
In Eq.~\ref{Eq:fractionAGB}, 
$M_{\rm AGB}$ corresponds to the mass of a halo star currently 
entering the TP-AGB phase. Before the TP-AGB phase, the star will have
gone successively through the main-sequence, the main
sequence turn-off (TO), the red-giant-branch phase (RGB), the core
helium-burning or horizontal-branch phase (HB), and the early-AGB
phase, the latter two being considered as a whole in what follows.    
Similarly to $M_{\rm AGB}$, one may define the threshold masses  $M_{\rm TO}$,
 $M_{\rm RGB}$, and $M_{\rm HB}$ for
stars formed at the birth of the Galaxy and 
currently entering these various evolutionary phases. 
These masses may be defined from evolutionary time-scale
considerations, since all of these stars are supposed to be co-eval with
the age of the Galaxy $\tau_{\rm Galaxy}$. Hence,
\begin{eqnarray*}
\tau_{\rm Galaxy} & = & \tau_{{\rm MS}}(0.8) + \tau_{{\rm TO}}(0.8) +
   \tau_{{\rm RGB}}(0.8) + \tau_{{\rm HB}}(0.8) + \tau_{{\rm AGB}}(0.8)\\
& = & \tau_{{\rm MS}}(M_{\rm AGB}) + \tau_{{\rm TO}}(M_{\rm AGB}) +
   \tau_{{\rm RGB}}(M_{\rm AGB}) + \tau_{{\rm HB}}(M_{\rm AGB}) \\
& = & \tau_{{\rm MS}}(M_{\rm HB}) + \tau_{{\rm TO}}(M_{\rm HB}) +
   \tau_{{\rm RGB}}(M_{\rm HB}) \\
& = & \tau_{{\rm MS}}(M_{\rm RGB}) + \tau_{{\rm TO}}(M_{\rm RGB}) \\
& = & \tau_{{\rm MS}}(M_{\rm TO}) 
\end{eqnarray*}   

The last equation of the above set indicates that the stars now at the
main-sequence TO have a mass $M_{\rm TO}$, so that their lifetime on
the main sequence equals the age of the Galaxy. The mass, $M_{\rm RGB}$,
of stars now entering the red-giant-branch stage is then given by combining
the last two equations in the above set:
\begin{equation}
\tau_{\rm MS}(M_{\rm TO}) - \tau_{\rm MS}(M_{\rm RGB}) = \tau_{\rm
     TO}(M_{\rm RGB}),  
\end{equation}
or, in other words, the difference $\delta\tau_{\rm MS}(M_{\rm RGB})$ 
of the main-sequence lifetimes of
the stars of masses  $M_{\rm TO}$ and $M_{\rm RGB}$ is just equal to
the time spent on the TO by the star of mass $M_{\rm RGB}$ currently
entering the RGB.  
This difference, $\delta\tau_{\rm MS}(M_{\rm RGB})$, relates to the mass
difference, $\delta M = M_{\rm TO} - M_{\rm RGB}$, through 
the lifetime, $\tau_{\rm MS}$, of main-sequence stars:
\begin{equation}
\tau_{\rm MS} \propto M^{-2.5},
\end{equation}
leading to
\begin{equation}
\label{Eq:deltatau}
\frac{\delta \tau_{\rm MS}}{\tau_{\rm MS}} = -2.5\; \frac{\delta M}{M}.
\end{equation}

\noindent Solving for $M_{\rm RGB}$ then yields:  
\begin{equation}
M_{\rm RGB}= M_{\rm TO}\; \left(1 + 0.4 \;
\frac{\tau_{\rm TO}(M_{\rm RGB})}{\tau_{\rm MS}(M_{\rm TO})}\right).
\end{equation}
The other masses are derived in a similar way:
\begin{eqnarray*}
M_{\rm HB} & = & M_{\rm TO}\; \left(1 + 0.4 \frac{\tau_{\rm TO}(M_{\rm
    HB})}{\tau_{\rm MS}(M_{\rm TO})} + 0.4 \frac{\tau_{\rm RGB}(M_{\rm HB})}{\tau_{\rm MS}(M_{\rm TO})}\right)\\
M_{\rm AGB} & = & M_{\rm TO}\; (1 + 0.4 \frac{\tau_{\rm TO}(M_{\rm
    AGB})}{\tau_{\rm MS}(M_{\rm TO})} + 0.4 \frac{\tau_{\rm
    RGB}(M_{\rm AGB})}{\tau_{\rm MS}(M_{\rm TO})}\\
    & & + 0.4 \frac{\tau_{\rm
    HB}(M_{\rm AGB})}{\tau_{\rm MS}(M_{\rm TO})})\\
0.8 & = & M_{\rm TO}\; (1 + 0.4 \frac{\tau_{\rm TO}(0.8)}{\tau_{\rm MS}(M_{\rm TO})} + 0.4 \frac{\tau_{\rm
    RGB}(0.8)}{\tau_{\rm MS}(M_{\rm TO})} \\
    & & + 0.4 \frac{\tau_{\rm
    HB}(0.8)}{\tau_{\rm MS}(M_{\rm TO})} + 0.4 \frac{\tau_{\rm
    AGB}(0.8)}{\tau_{\rm MS}(M_{\rm TO})})
\end{eqnarray*}

The last equation above is used to derive $M_{\rm TO}$. An iterative procedure is
required, since the lifetime ratios must be normalized by the main-sequence
lifetime of a star of yet-unknown mass, $M_{\rm TO}$. In the first iteration, the various
lifetime ratios {\it normalized by the main-sequence lifetime of the 0.8}~\Msun\ star
(namely $12.2\;10^9$~y, up to the turn-off at $M_{\rm bol} \sim 3.5$) are read from
Fig.~\ref{Fig:age0.8}, namely
$\tau_{\rm
    HB}(0.8) / \tau_{\rm MS}(0.8) = 0.006$, $\tau_{\rm
    RGB}(0.8) / \tau_{\rm MS}(0.8) = 0.045$, and 
$\tau_{\rm TO}(0.8) / \tau_{\rm MS}(0.8) = 0.08$.

The lifetime spent on the TP-AGB, $\tau_{\rm AGB}$, is very sensitive to the
adopted mass-loss rate prescription. Very little is known about mass loss in
extremely low-metallicity stars \citep{VanLoon-2006}. If it is driven by
radiation pressure, then stellar wind models \citep{Kudri-1987,Kudri-1991,
VanLoon-00} indicate that the rate depends roughly on the square root of the
metallicity. Observational support for the sensitivity of mass-loss rate to
metallicity has been provided by \citet{Groenewegen-1995}, who compare the dust
content of LMC and galactic AGB stars of similar pulsational periods.
Alternatively, \citet{Vassiliadis-Wood-1993} derived an empirical formulation of
the mass-loss rate based on the pulsational properties of the Mira variables and
pulsating OH/IR stars. Their expression is not explicitly dependent on
metallicity, but it does include the effect of the chemical composition through
the stellar parameters entering the computation of the pulsational period. When
the mass-loss rate depends upon metallicity, it remains quite low (and hence the
AGB lifetime quite long) unless third dredge-up (3DUP) episodes increase the
metal content of the envelope and boost the mass-loss rate. This prescription
implies a longer lifetime on the TP-AGB than the Vassiliadis \& Wood
prescription. Therefore, the TP-AGB lifetime displayed in Figs.~\ref{Fig:ageLionel} and \ref{Fig:age0.8}, computed with the Vassiliadis \& Wood
prescription, is likely to represent a lower limit on the actual lifetime.
Moreover, in our calculations, only 26 pulses have taken place, and the envelope
mass still amounts to 0.1~\Msun\ with a mass loss rate of $4\;
10^{-8}$~\Msun~y$^{-1}$. The TP-AGB lifetime displayed in Fig.~\ref{Fig:ageLionel} for the 0.8~\Msun\ model therefore represents a lower bound to the
actual lifetime, which may be expected to be longer by a few $10^6$~y.

  \begin{figure}
    \centering \includegraphics[width=8cm]{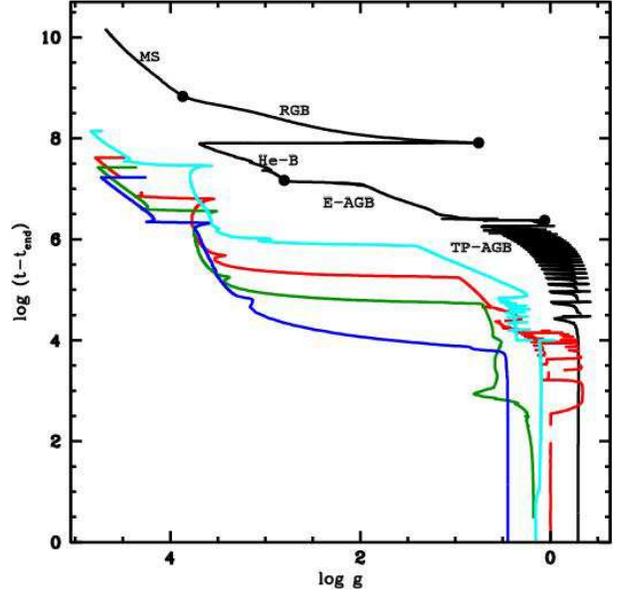}
       \caption{Same as Fig.~\ref{Fig:HRLionel} for the ages of the five
models as a function of the effective gravity. The age is counted backward
from the last computed model. The occurrence of thermal pulses is evident
from the rapid oscillations of log~$g$ near the end of the
evolution. The large dots on the 0.8~\Msun\ track mark the transition
between different evolutionary phases: main sequence (labelled MS),
red giant branch (RGB), core helium-burning (He-B), early AGB (E-AGB),
and thermally-pulsing AGB (TP-AGB).}
          \label{Fig:ageLionel}
    \end{figure}

\begin{figure}
   \centering
   \includegraphics[width=8cm]{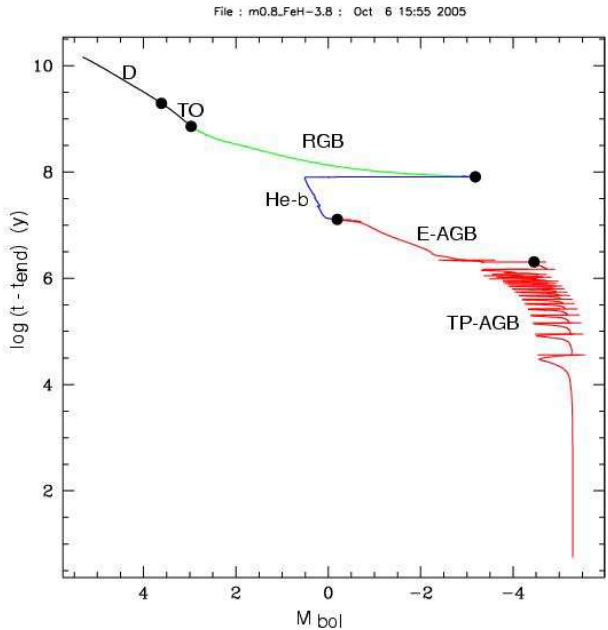}
      \caption{Same as Fig.~\protect\ref{Fig:ageLionel} for the
              0.8~\Msun\ model, but expressed as a function of $M_{\rm
      bol}$. The lifetimes in the different evolutionary stages are
      used in Sect.~\ref{Sect:proba}. Note that the AGB lifetime
      provided by this figure is a lower limit to the actual value,
      since only 26 pulses have been computed, and the remaining
      envelope mass amounts to 0.1~\Msun. A few million years should thus be
added to the  AGB
    lifetime displayed on the figure. Moreover, the Vassiliadis \& Wood (1993)
      prescription for the mass loss used in these computations (see
      text) also shortens the AGB lifetime.   
              }
         \label{Fig:age0.8}
   \end{figure}

Given the uncertainty affecting $\tau_{\rm AGB}$, it will be kept as a free
parameter for the remainder of this section. Nevertheless, as an illustrative
example, conservatively adopting $\tau_{\rm AGB}(0.8) = 10^7$~y yields
$\tau_{\rm AGB}(0.8) / \tau_{\rm MS}(0.8) = 8\; 10^{-4}$, hence $\tilde{M}_{\rm
TO} = 0.760$~\Msun. From Eq.~\ref{Eq:deltatau}, such a star has a main-sequence
lifetime $1.5\;10^9$~y longer than the 0.8~\Msun\ one, or $\tau_{\rm MS}(M_{\rm
TO}) / \tau_{\rm MS}(0.8) = 1.12$. This factor is thus applied to the above
lifetime ratios to normalize them by $\tau_{\rm MS}(M_{\rm TO})$, as required.
The adopted ratios are listed in Table~\ref{Tab:distcutoff}. This second (and
last) iteration then yields $M_{\rm TO} = 0.764$~\Msun. Since all five threshold
masses are very close to one another, the normalized lifetimes appearing in the
equation set above will all be taken equal to the values corresponding to the
0.8~\Msun\ model listed in Table~\ref{Tab:distcutoff}.

To describe a real stellar sample (which in this case is magnitude limited), the
observational censorship needs to be included, implying that the sampling volume
will depend on the luminosity of the various classes of stars, thus largely
favouring bright stars, such as AGB stars, over main-sequence stars. In the
following, the integral appearing in the denominator of Eq.~\ref{Eq:fractionAGB}
will be split into five different components -- thermally-pulsing AGB (TP-AGB)
stars, horizontal-branch and early AGB stars (collectively referred to as HB),
main-sequence turnoff stars (TO), and finally main-sequence stars (MS). The
lower limit of the main-sequence mass range considered in Eq.~\ref{Eq:fractionAGB} is denoted $M_{\rm low}$; however, as shown below, it has
no significant impact on the final result.

With these conventions, Eq.~\ref{Eq:fractionAGB} can be rewritten
\begin{equation}
\label{Eq:Proba}
P_{\rm AGB} = \frac{n_{\rm vol_{\rm AGB}}
   \int_{M_{\rm AGB}}^{0.8 M_\odot} \Psi_{\rm IMF} (M)\; {\rm d}  M}
{\Sigma_{i=1}^{6} n_{{\rm vol},i} \int_{M_i}^{M_{i+1}} \;
\Psi_{IMF} (M)\; {\rm d} M},
\end{equation}
where $n_{\rm vol}$ is the number of halo stars in the volume sampled for the
various subtypes of stars (namely TP-AGB, HB, RGB, TO, and MS; $M_1 = M_{\rm
low}, M_2 = M_{\rm TO}$..., $M_6 = 0.8$~\Msun), given the fact that the HK
sample is limited to the $11 \le B \le 16$ magnitude range \citep{Beers-1992}.
The adopted volume is a cone along the line of sight to CS~30322-023, truncated
at a minimum distance fixed by $B=11$, and the absolute $B$ magnitude of the
considered stellar subtype, and at a maximum distance similarly fixed by the
condition $B=16$. The halo star number density is assumed to decrease as
$[R^2+(z/0.76)^2]^{-2.44/2}$, $R$ being the galactocentric distance measured in
the galactic plane and $z$ the distance from the plane, according to the model
of \citet{Robin-2003}.

The distance cut-offs, $d_{\rm min, max}$, for the RGB, HB, and AGB stars are
listed in Table~\ref{Tab:distcutoff}, along with the adopted bolometric
magnitudes and effective temperatures (from the \scshape STAREVOL \upshape
tracks discussed in Sect.~\ref{Sect:evol} and displayed in Fig.~\ref{Fig:HR-0.8}), bolometric corrections and total galactic absorption in the $B$ band
along the line of sight to CS~30322-023, according to the discussion in
Sect.~\ref{Sect:parameters}. For TO and MS stars, galactic absorption is assumed
to be negligible, since the volume sampled is small. The bolometric correction
in the $B$ band is taken from a grid of MARCS models with [Fe/H]~$=-4$, and a
range of gravities and effective temperatures. They do not differ much from
those provided in Table~6 of \citet{Alonso-1999} for giants with [Fe/H]~$=-3$.
The bolometric correction for AGB stars has been taken identical to that derived
for \CS\ (Sect.~\ref{Sect:observations}). 

Main-sequence stars are sampled in an exceedingly small volume as compared to
the other, more luminous classes, as may be judged from Table~\ref{Tab:distcutoff}. Those stars should thus be very rare among the HK survey
\citep{Beers-85} since the volume effect largely offsets the effect of the
increasing IMF, hence the first term ($i=1$) of the sum in Eq.~\ref{Eq:Proba}
is neglected in the following.

\begin{table*}
\caption[]{
Sampling distances of the HK survey for the various stellar
subtypes and fractional lifetimes $\tau (0.8)/\tau_{\rm MS}(M_{\rm TO})$ for the
various evolutionary phases of the 0.8~\Msun\ model of [Fe/H]$ = -3.8$. The lifetime
in the AGB phase is subject to various uncertainties, and has therefore not been
listed. $BC_B$ and $A_B$ are, respectively, the bolometric correction and
total absorption in the $B$ band.
\label{Tab:distcutoff}
}
\begin{tabular}{llllllllllll}
\hline
\hline
type & $\tau_i(0.8)/\tau_{\rm MS}(M_{\rm TO})$ & $\log T_{\rm eff}$
& $M_{\rm bol}$ & $BC_B$ & $A_B$ &
$d_{\rm min}$ & $d_{\rm max}$ & $n_{\rm  vol}/n_{\rm vol,TO}$ & \\
$i$     & & &  &&& \multicolumn{1}{c}{(kpc)}   & \multicolumn{1}{c}{(kpc)} \\ 
\hline
AGB & -                      & 3.61 & $-$5  &$-$1.6 & 0.2 & 6.91 & 69.1
&274\\ 
HB + E-AGB & $5.6\; 10^{-3}$ & 3.85 & 0.5 & $-$0.5 & 0.2 & 0.91 &
9.12 & 40.8\\ 
RGB & $4.0\; 10^{-2}$        & 3.71 & 0.5 & $-$1.0 & 0.2 & 0.72 & 7.24 &
24.7\\ 
TO & $6.9\; 10^{-2}$         & 3.85 & 3.5 & $-$0.5 & 0.0 & 0.25 & 2.51 & 1\\
MS  & 0.88                   & 3.81 & 5   & $-$0.5 & 0.0 & 0.12 & 1.23 & -\\
\hline
\end{tabular}
\end{table*}

The number of stars in  a truncated cone along the line of sight to CS~30322-023, with an opening angle of $\alpha$ (this parameter, appearing in both the numerator and denominator of Eq.~\ref{Eq:Proba}, has no impact on the final result) thus becomes
\begin{equation}
\label{Eq:nvol}
n_{\rm vol}(M) = \pi\; {\rm tg}^2\alpha \; \rho_0 \int_{d_{\rm min}(M)}^{d_{\rm max}(M)} \left(R^2 + (z/0.76)^2\right)^{-1.22} \; d^2\; {\rm d}\,d
\end{equation}
 where $R = r_0 - d \cos b$ (or $d \cos b - r_0$, whichever is positive) 
is the galactocentric distance measured in the galactic 
plane, $z = d \sin b$ the distance from the plane, 
$b = -46.3^\circ$ is the galactic latitude of CS~30322-023 
(its galactic longitude, $l = -6.8^\circ$, is sufficiently close to 0 to be
 neglected in the above formula), 
$r_0 \sim 8$~kpc is the galactocentric distance of the Sun, and
$\rho_0$ is the central galactic density of extremely metal-poor halo
stars. The latter parameter will later cancel, and thus has no impact
on the final result.

The integrals of Eq.~\ref{Eq:nvol} have been computed numerically; the results
are listed in Table~\ref{Tab:distcutoff}. The final proportion of AGB stars in
the co-eval sample is displayed in Fig.~\ref{Fig:fractionAGB}. Given the fact
that a reasonable estimate for the fractional TP-AGB lifetime, $\tau_{\rm
AGB}(0.8) / \tau_{\rm MS}(M_{\rm TO})$, is on the order of $10^{-3} (=
10^7/10^{10}$~y/y), the main conclusion from the present analysis is that it is
{\it not at all unlikely} to observe an extremely low-metallicity TP-AGB star in
a co-eval sample. The \citet{Beers-1992} sample is most likely not a co-eval
sample, as it includes stars with metallicities larger than that of \CS\ , which
must have formed somewhat more recently. In a sample containing stars formed
over a long period of time (with a constant star formation rate), the relative
fractions of stars of various kinds are proportional to their lifetimes. Since
the lifetime of stars at the main-sequence turnoff is about 70 times longer than
the AGB lifetime, it is not surprising that \citet{Beers-1992} sample contains a
much larger fraction of TO stars than predicted under the co-evality hypothesis.
Nevertheless, it should be stressed that the above analysis was not designed to
mimick the \citet{Beers-1992} sample (this would require a detailed knowledge of
the star formation history in the halo, which is beyond the scope of the present
paper), but rather to roughly evaluate the likelihood of finding an extremely
low-metallicity TP-AGB star. Incidentally, an analysis similar to the one
presented here, based on the Bahcall-Soneira Galaxy model \citep{Bahcall-Soneira-1980},
was presented by \citet{Beers-85}, who arrived at the same conclusion -- a
magnitude-limited sample of halo stars to apparent magnitude $B = 16$ is,
by far, dominated by giant stars.  

\begin{figure}
   \centering
   \includegraphics[width=8cm]{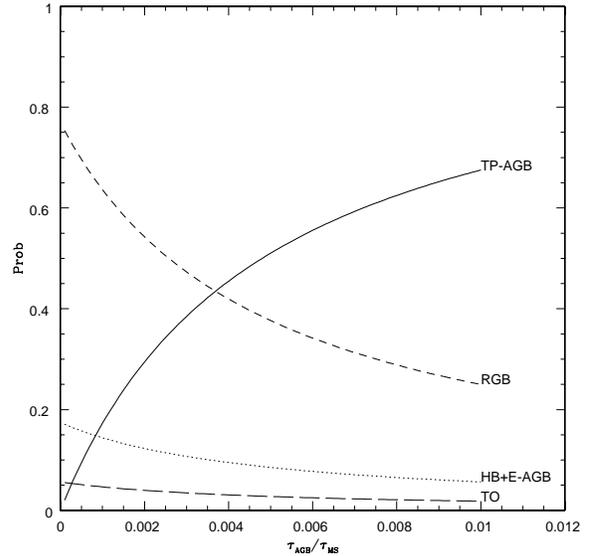}
      \caption{Fraction of AGB stars (solid line), HB
              stars (dotted line), RGB stars (short-dashed line), and TO stars
(long-dashed line)  in a co-eval sample born in the early Galaxy. 
              }
         \label{Fig:fractionAGB}
   \end{figure}

\section{Abundances}
\label{Sect:abundances}

\begin{figure}
   \centering
   \includegraphics[height=0.5 \textwidth,angle=-90]{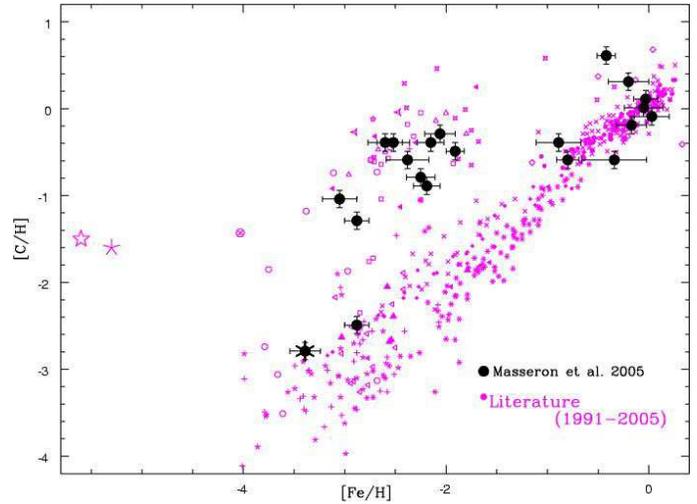}
      \caption{[C/H] as a function of metallicity for metal-poor stars,
	collected from the literature \protect\citep[see ][ for
	detailed references]{Masseron-2006:b}. The lower diagonal sequence
corresponds to the trend expected from  galactic chemical evolution
	for unpolluted stars. Two of the stars from the sample of 
\citet{Masseron-2006:b}, among which are \CS\ (asterisk), 
fall on this galactic
	sequence. The picture
changes, however, when considering C+N (Fig.~\protect\ref{Fig:CNFeH}). 
%The carbon abundance predicted by the
%0.8~\Msun\ model of metallicity [Fe/H] = $-3.8$ is represented by the
%pentagon. It overestimates the carbon abundance in all carbon-rich
%metal-poor stars.  
              }
         \label{Fig:CvsFe}
   \end{figure}

\begin{figure}
   \centering
   \includegraphics[width=8cm]{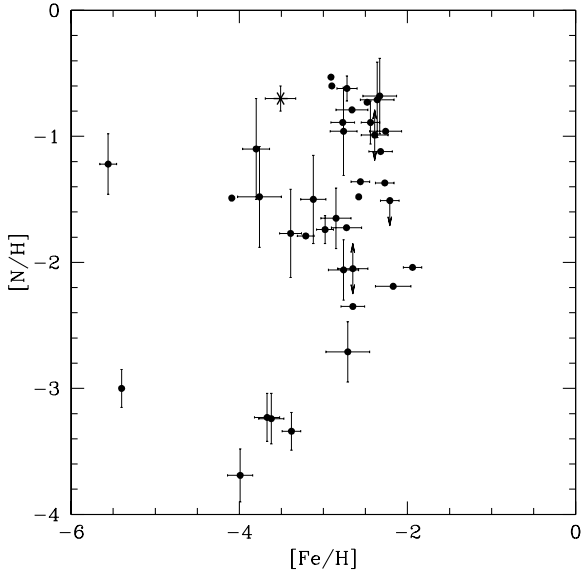}
      \caption{
Same as Fig.~\protect\ref{Fig:CvsFe} for [N/H] as a 
function of metallicity for CEMP stars
collected from the literature (see Sect.~\ref{Sect:evol} for
the full reference list). As usual, \CS\ is represented by an asterisk. 
%Same as Fig.~\protect\ref{Fig:CvsFe} for [N/Fe] as a 
%function of metallicity for CEMP stars
%	collected from the literature (see Sect.~\ref{Sect:evol} for
%	the full 
%	reference list). The dashed line corresponds to [N/H]$ = -0.5$. As
%	usual, \CS\ is represented by an asterisk. 
              }
         \label{Fig:NvsFe}
   \end{figure}

\begin{figure}
  \centering \includegraphics[width = 0.5 \textwidth]{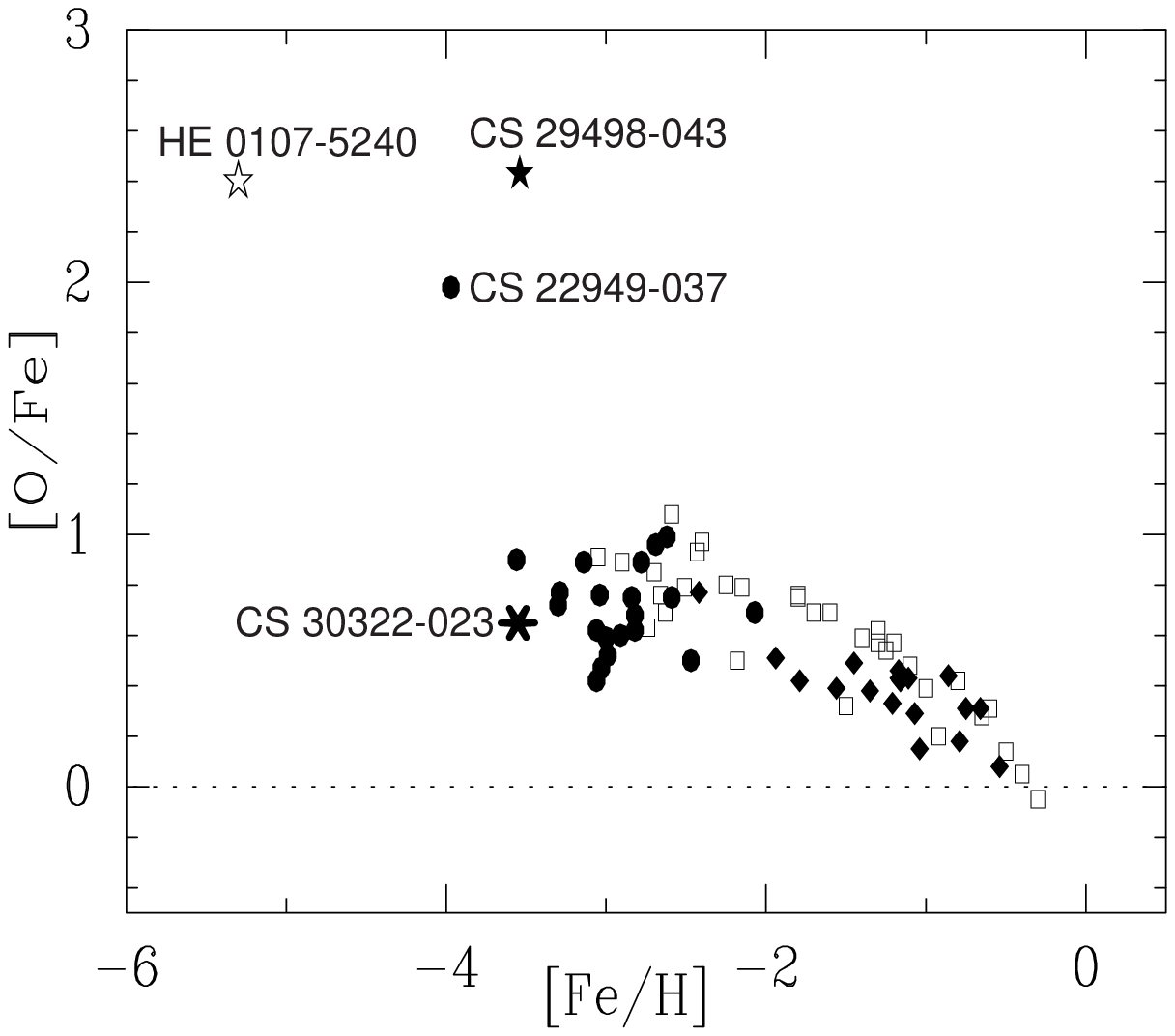}
\caption[]{
[O/Fe] as a function of [Fe/H] for \protect\CS\ (asterisk), CS~29498-043
 \citep{Aoki-2004}, HE~0107-5240 \citep{Bessell-2004} and
for unevolved dwarf stars from \protect\citet{Israelian-2001} (open
squares),
\citet{Cayrel-2004} (filled circles), and \protect\citet{Nissen-2002}
(filled diamonds). The filled symbols correspond to [O/Fe] values derived from
 the [O I] line, while open symbols correspond to oxygen abundances derived 
from the triplet or OH lines. No 3D corrections have been applied. 
Figure adapted from \citet{Aoki-2004}.
}
  \label{Fig:OFe}
\end{figure}

\begin{figure}
   \centering
   \includegraphics[width=8cm]{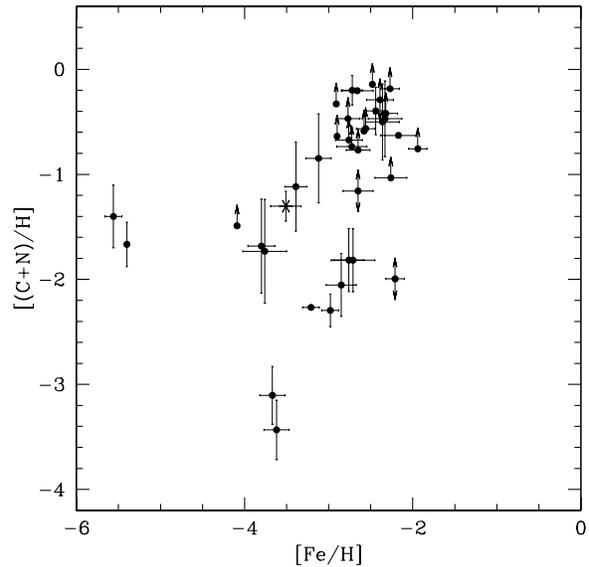}
      \caption{ [(C+N)/H]
      as a function of metallicity for the same sample of
      CEMP stars as in Fig.~\protect\ref{Fig:NvsFe}.}
         \label{Fig:CNFeH}
   \end{figure}

\renewcommand{\baselinestretch}{ 0.9}
\begin{table}
\caption{Abundances for CS~30322-023 obtained with the plane-parallel
  model. $N$ is the number of lines used in the analysis; `syn'
  indicates that spectral synthesis has been used to derive the
  abundance. The column
  $\log \epsilon ({\rm X})$ lists the abundance in the scale where 
$\log \epsilon({\rm H}) = 12$. Abundances are normalized by the solar data 
from \protect\citet{Asplund-05}.}
\label{tab:abund}
\begin{tabular}{l|c|c|c|c|c} 
\hline
Element  & $N$ & $\log \epsilon({\rm X})$ & $\sigma$  & [X/H] & [X/Fe]    \\
\hline
C (CH)&  (syn)  &      5.60 &  -      &  -2.79  &   0.60  \\
N (CN)&  (syn)  &      7.20 &  -      &  -0.58  &   2.81  \\
 O  I &  1(syn) &      5.90 &  -      &  -2.76  &   0.63  \\
Na  I &  4      &      4.07 &  0.24   &  -2.10  &   1.29  \\
Mg  I &  4      &      4.94 &  0.10   &  -2.59  &   0.80  \\
Al  I &  2      &      3.70 &  0.3    &  -2.67  &   0.72  \\
Si  I &  1      &      4.70 &  0.2    &  -2.81  &   0.58  \\
 K  I &  1(syn) &   $<$3.00 & -       &  -2.08  &   1.31  \\
Ca  I & 10      &      3.37 &  0.12   &  -2.94  &   0.45  \\
Sc  II&  4      &     -0.12 &  0.17   &  -3.17  &   0.24  \\
Ti  I & 10      &      1.31 &  0.19   &  -3.59  &  -0.20  \\
Ti  II& 14      &      1.72 &  0.13   &  -3.18  &   0.23  \\
 V  II&  3(syn) &      0.40 &  0.1    &  -3.60  &  -0.19  \\
Cr  I &  9      &      1.66 &  0.22   &  -3.98  &  -0.59  \\
Mn  I &  6      &      1.66 &  0.28   &  -3.73  &  -0.34  \\
Fe  I & 66      &      4.06 &  0.18   &  -3.39  &   0     \\
Fe  II&  7      &      4.04 &  0.19   &  -3.41  &   0     \\
Co  I &  1      &      1.33 &  -      &  -3.59  &  -0.20  \\
Ni  I &  6      &      2.56 &  0.18   &  -3.67  &  -0.28  \\
Cu  I &  1(syn) &     -0.10 &  -      &  -4.31  &  -0.92  \\
Zn  I &  2      &      1.23 &  0.04   &  -3.37  &   0.02  \\
Sr  II&  1(syn) &     -1.00 &  -      &  -3.92  &  -0.51  \\
 Y  II&  6      &     -1.56 &  0.15   &  -3.77  &  -0.36  \\
Zr  II&  5      &     -0.68 &  0.30   &  -3.27  &   0.14  \\
Tc  I &  1(syn) &  $<$-1.40 & -       &  -      &   -     \\
Ru  I &  1(syn) &  $<$-0.50 & -       &  -2.34  &   1.05  \\
Rh  I &  1(syn) &  $<$-2.00 & -       &  -3.12  &   0.27  \\
Ba  II&  3(syn) &     -0.70 &  0.1    &  -2.87  &   0.54  \\
La  II&  8(syn) &     -1.80 &  0.1    &  -2.93  &   0.48  \\
Ce  II& 18      &     -1.22 &  0.24   &  -2.80  &   0.61  \\
Pr  II&  3(syn) &     -2.50 &  0.10   &  -3.21  &   0.20  \\
Nd  II& 24      &     -1.34 &  0.29   &  -2.79  &   0.62  \\
Sm  II&  4(syn) &     -2.30 &  0.2    &  -3.31  &   0.10  \\
Eu  II&  1(syn) &     -3.40 &  -      &  -3.92  &  -0.51  \\
Gd  II&  1(syn) &  $<$-2.30 &  -      &  -3.42  &  -0.01  \\
Tb  II&  1(syn) &  $<$-3.20 &  -      &  -3.48  &  -0.07  \\
Dy  II&  2(syn) &     -2.40 &  0.1    &  -3.54  &  -0.13  \\
Er  II&  1(syn) &     -2.56 &  -      &  -3.49  &  -0.08  \\
Tm  II&  1(syn) &  $<$-3.50 &  -      &  -3.50  &  -0.09  \\
Yb  II&  1      &     -2.10 &  -      &  -3.18  &   0.23  \\
Hf  II&  2(syn) &     -2.00 &  0.1    &  -2.88  &   0.53  \\
W   I &  1(syn) &  $<$-1.50 &  -      &	 -2.61 	&   0.78 \\
Pb  I &  2(syn) &      0.10 &  0.2    &  -1.90  &   1.49  \\
\hline
\end{tabular}
\vspace*{1 mm}\\
$^{12}$C/$^{13}$C = $4 \pm 1$ \\
\end{table}

\begin{figure}
\resizebox{0.45 \textwidth}{!}{\includegraphics{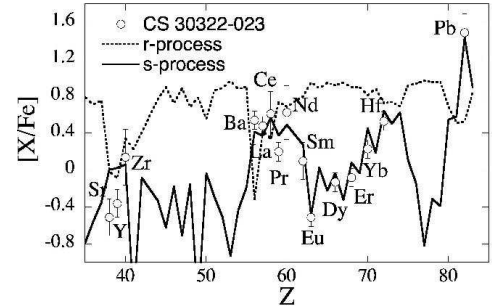}}
\caption{Abundance pattern of CS~30322-23 compared to two model
  predictions for the s- and r-processes.
The s-process abundance pattern has been computed by
  post-processing an AGB model of mass 0.8~\Msun\ and metallicity
  [Fe/H]~$=-3.8$, as described in \cite{Goriely-00}. The r-process pattern corresponds to the scaled solar r-process abundances, from \citet{Palme-Beer-1994}.   
}\label{fig:abund_pattern}
\end{figure}

Because of the general crowding of the spectra for cool, evolved stars with
molecular lines, the analysis of stars as \CS\ is very challenging. Thanks to
the low metallicity, however, abundances could be obtained for 35 species, while
seven received upper limits on their abundances with the plane-parallel (PP) model.
These results are summarised in Table~\ref{tab:abund}. Hyper-fine structure has
been included for Ba, La, Pr, Eu, Tb, Dy, Er, Lu, and Pb. C and N abundances are
derived from the CH and CN lines. The O abundance is derived from the 630.0~nm
[O I] forbidden line which, among the O abundance diagnostics, is expected to be
the least affected one by non-LTE conditions \citep{Israelian-2004}, although it
is sensitive to 3D granulation effects \citep{Shchukina-2005}. 

Although the gravity has been derived from the Fe~I/Fe~II ionisation balance, a
large discrepancy (about 0.4~dex) is nevertheless found between the abundances
derived from the Ti~I and Ti~II lines. This discrepancy might be ascribed to
various causes: (i) the use of a plane-parallel instead of a spherical model
(but see the discussion in Sect.~\ref{Sect:parameters}), (ii) NLTE effects 
\citep{Thevenin-99,Israelian-2001,Israelian-2004,Korn-2003}, 
(iii) 3D effects \citep{Asplund-2005}, or (iv) a systematic shift between the
scales on which the $\log gf$ values for Ti~I and Ti~II are derived. A similar
problem has been reported for Cr in other cool metal-poor stars
\citep{Johnson-2002a}. The first three possibilities should be investigated in
the future by using tailored 3D, NLTE models.

The uncertainty quoted in Table~\ref{tab:abund} in the column labelled $\sigma$
does not include such possible systematic errors, as it only refers to the
standard deviation of the abundances derived from the different lines used for a
given species. When only one line is available, spectrum synthesis has been
used, and a 0.1~dex uncertainty should in this case be considered as typical,
arising from errors associated with the location of the continuum and on the
line-fitting process.
 
Table~\ref{tab:abund} reveals that, quite remarkably, the C enhancement in \CS\
is only moderate, in contrast to the N enhancement (Figs.~\ref{Fig:CvsFe} and
\ref{Fig:NvsFe}). Na is strongly enhanced as well. The C/O ratio is only 0.50.
The C and O abundances are in fact compatible with the trend due to the chemical
evolution of the Galaxy, as displayed in Fig.~\ref{Fig:CvsFe} for C and in
Fig.~\ref{Fig:OFe} for O. However, because the carbon isotopic ratio is low
($^{12}$C/$^{13}$C = 4), it is unlikely that C has retained its initial value in
the envelope of \CS. As further discussed in Sect.~\ref{Sect:discussion}, the
combination of a low $^{12}$C/$^{13}$C ratio and large N overabundance strongly
suggests that pollution of the atmosphere by material processed by the CN cycle
has occurred. In this respect, it is worth noting that \CS\ falls among the
other CEMP stars in a diagram showing [(C+N)/H] as a function of metallicity
(Fig.~\ref{Fig:CNFeH}). The $\alpha$-elements in \CS\ do not exhibit any
overabundances with respect to the values expected for its metallicity from
galactic chemical evolution. There is also no observed enhancement of r-process
elements, whereas the s-process elements belonging to the second and third peaks
are clearly enhanced (Fig.~\ref{fig:abund_pattern}). Tc~I and Tc~II lines have
been searched for, and not found. As shown on Fig.~\ref{Fig:Tc}, an upper limit
$\log \epsilon({\rm Tc}) < -1.4$ (translating to $\epsilon({\rm Tc})
/\epsilon({\rm Zr}) < 0.2$) may be derived from the Tc~I~$\lambda 426.23$~nm
line. This upper limit on the Tc abundance is not meaningful, however, since
$\epsilon({\rm Tc})/\epsilon({\rm Zr}) \sim 0.001$ is expected from s-process
nucleosynthesis \citep[see Fig.~15 of ][]{Goriely-00}. The difficulty of
detecting the Tc~I line in this star results from its rather high temperature
(as compared to solar-metallicity AGB stars, which are much cooler), implying
that Tc is mostly ionized. Unfortunately, the strongest Tc~II lines lie in the
far ultraviolet; perhaps these lines could be searched for from space in the
future.

\begin{figure}
   \centering \includegraphics[width=8cm,height=8cm]{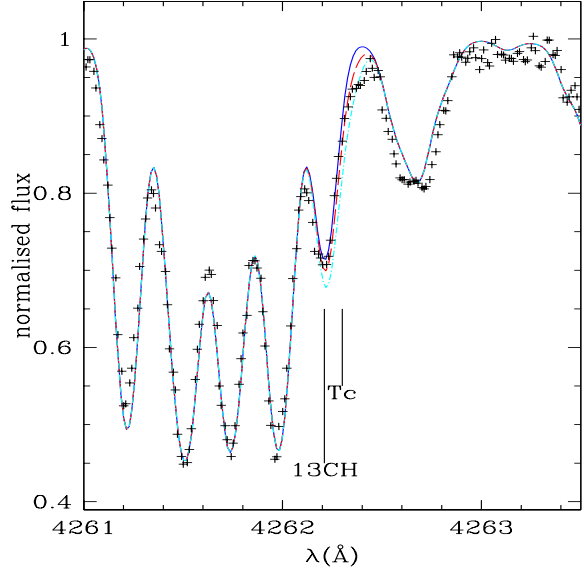}
      \caption{The observed spectrum of \protect\CS\ (crosses) 
around the Tc~I~$\lambda
426.23$~nm line, with three synthetic spectra: without Tc (blue solid
line), with $\log
\epsilon({\rm Tc}) = -1.4$ (red dashed line), and with $\log
\epsilon({\rm Tc}) = -1.0$ (cyan dot-dashed line). 
%The poor sensitivity of the Tc line to its abundance is due due to
%blend with a strong $^{13}$CH line.
}
         \label{Fig:Tc}
   \end{figure}

The abundance pattern for heavy elements is presented in 
Fig.~\ref{fig:abund_pattern}, along with the scaled solar r-process abundance pattern, and an
s-process abundance pattern expected from proton-mixing, forming a $^{13}$C
pocket below the He-H discontinuity in a 0.8~\Msun\ AGB star of metallicity
[Fe/H]~$=-3.8$ \citep{Goriely-00}. With the exception of Sr and Y, all observed
abundances almost perfectly match the predictions of a pure s-process operating
at low metallicity in the framework of the proton-mixing scenario
\citep{Straniero-95,Goriely-00}. The s-process abundance pattern in \CS\ is at
the lowest metallicity observed to date. As expected, it differs from its
solar-metallicity counterpart due to the large Pb overabundance 
\citep{Goriely-00,Busso-01,Goriely-Siess-01}, which matches the
value observed in \CS\ (Figs.~\ref{fig:abund_pattern} and \ref{Fig:PbBaFeH}),
thus making this star a new member of the class of lead stars \citep{VanEck-01,
VanEck-03}. The large spread in the Pb abundances observed among CEMP stars
(Fig.~\ref{Fig:PbBaFeH}) is surprising, however, as is further discussed in
Sect.~\ref{Sect:discussion}. The parameter responsible for this scatter remains
unidentified, as the scatter does not correlate with either [C/Fe] or N/(C+N). 

\begin{figure}
   \centering \includegraphics[width=8cm,height=8cm]{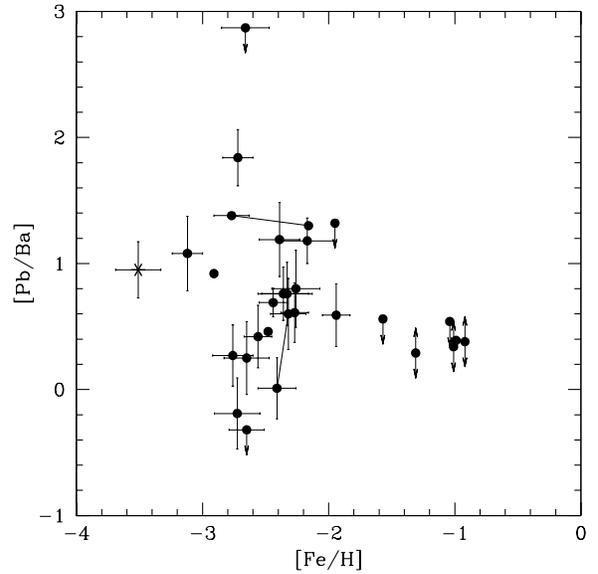}
      \caption{The [Pb/Ba] ratio as a function of metallicity for
	the sample of CEMP stars (excluding r-process-enhanced stars) 
taken from the literature (see the references listed in 
Sect.~\protect\ref{Sect:evol}). As before, \protect\CS\ is represented
	by an asterisk.}
         \label{Fig:PbBaFeH}
   \end{figure}

\begin{figure}
   \centering
   \includegraphics[width=8cm]{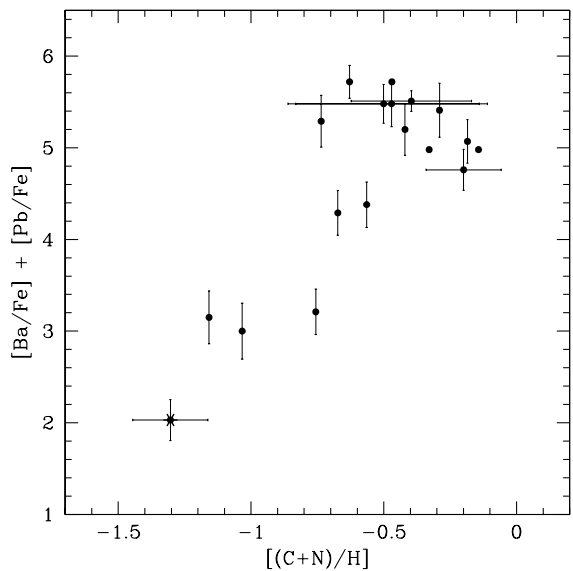}
      \caption{The global overabundance of s-process elements in CEMP
	stars (excluding r-process-enhanced stars), expressed in terms of
[Ba/Fe] + [Pb/Fe], as a function of [(C+N)/H]. Although the general
	trend may be explained by the dilution of the processed
	material in the envelope, there remains a significant --
	unexplained -- scatter in the s-process abundances 
(at a given [(C+N)/H]).
              }
         \label{Fig:PbBaCN2}
   \end{figure}

\begin{figure}
   \centering
   \includegraphics[width=8cm]{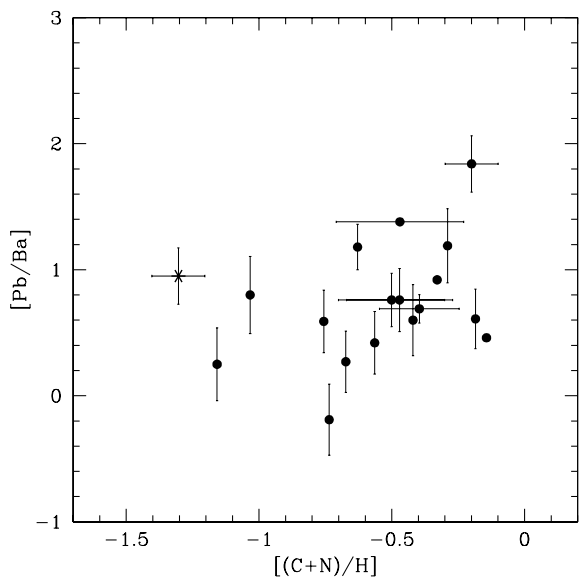}
      \caption{Same as Fig.~\protect\ref{Fig:PbBaFeH} for [Pb/Ba] as a 
function of [(C+N)/H].
}
\label{Fig:PbBaCNH}
\end{figure}

%\begin{figure}
%   \centering \includegraphics[width=8cm,height=8cm]{fig16a.ps}
%   \centering \includegraphics[width=8cm,height=8cm]{fig16b.ps}
%   \centering \includegraphics[width=8cm,height=8cm]{fig16c.ps}
%      \caption{Same as Fig.~\protect\ref{Fig:PbBaFeH} for Pb/Ba as a
%	function of [C/Fe] (upper panel) and [C+N/H] (lower panel). 
%	}
%         \label{Fig:PbBaCFe}
%   \end{figure}

To estimate the global overabundance of s-process elements in CEMP stars -- when
evaluating, for instance, its possible correlation with the C and N abundances
-- it is important to consider not only overabundances of elements (such as Ba)
belonging to the second s-process peak, but also overabundances of elements
(such as Pb) from the third peak. Failure to do so would largely underestimate the
s-process overabundance whenever lead stars are involved. This is very clear
from Fig.~\ref{fig:abund_pattern}. Fig.~\ref{Fig:PbBaCN2} reveals that there is
indeed a general tendency for ([Ba/Fe] + [Pb/Fe]) to increase with [(C+N)/H]
(this trend may in fact be ascribed to the dilution of the processed material in
a more or less massive envelope), although there remains a considerable scatter.
On the contrary, [(C+N)/H] exhibits little correlation with the s-process
efficiency expressed in terms of [Pb/Ba] (Fig.~\ref{Fig:PbBaCNH}).  

In absolute terms, the s-process overabundance ([Ba/Fe] + [Pb/Fe]) in \CS\ is
thus moderate, when compared to other CEMP stars (Fig.~\ref{Fig:PbBaCN2}), which
is suggestive of a large dilution factor of the processed material in the
envelope. 

Considering the sum Ba+Pb, instead of solely Ba, is also of interest when
discussing the [Ba/Eu] ratio. Fig.~\ref{Fig:EuBa} shows that, in a [Ba/Eu] vs.
[C/Fe] diagram, there are several stars with [Ba/Eu] ratios that are
intermediate between the solar s-process and solar r-process ratios. These stars
have been coined `s+r' or `s/r' stars \citep{Hill-2000,Cohen-2003a,
Sivarani-2004,Barbuy-2005,Ivans-2005,Jonsell-2006}; their [Ba/Eu] ratios do
indeed seem peculiar when compared to the solar s-process ratio 
(Fig.~\ref{Fig:EuBa}). Detailed s-process computations in low-metallicity AGB stars 
\citep{Goriely-Siess-2006} have shown that the [Ba/Eu] ratio predicted under
such circumstances are quite different from the solar s-process Ba/Eu
ratio (upper panel of Fig.~\ref{Fig:EuBa}), and matches at least some of the
observed [Ba/Eu] values. When considering the [(Ba+Pb)/Eu] ratio, instead of the
[Ba/Eu] ratio, s-process predictions become more robust (in the sense that the
different environments yield similar values, within 0.6~dex). Such predictions
may account for some of the stars displayed in the lower panel of 
Fig.~\ref{Fig:EuBa}, but the need for a separate class of `s+r' (`s/r') stars is even
stronger, as they separate even more clearly from pure s- and pure r-stars when
considering [(Ba+Pb)/Eu] instead of [Ba/Eu].

%But when compared to the
%predictions of an s-process operating in thermal pulses of massive,
%low-metallicity AGB stars, the Ba/Eu ratios observed in s+r stars do
%not look so peculiar any longer \citep{Goriely-Siess-2006}. One may
%even wonder whether the scatter observed among CEMP stars in the upper
%panel of Fig.~\ref{Fig:EuBa} could not be attributed to the fact that
%Ba is no longer a good indicator of the efficiency of the s-process in
%low-metallicity stars, and should be replaced by the sum Ba + Pb
%instead, as done in the lower panel of Fig.~\ref{Fig:EuBa}.
%Considering ([Ba/Eu] + [Pb/Eu]) instead of [Ba/Eu] indeed makes the
%need for a separate class of 's+r' stars much less clear.

\begin{figure}
   \centering
   \includegraphics[width=8cm]{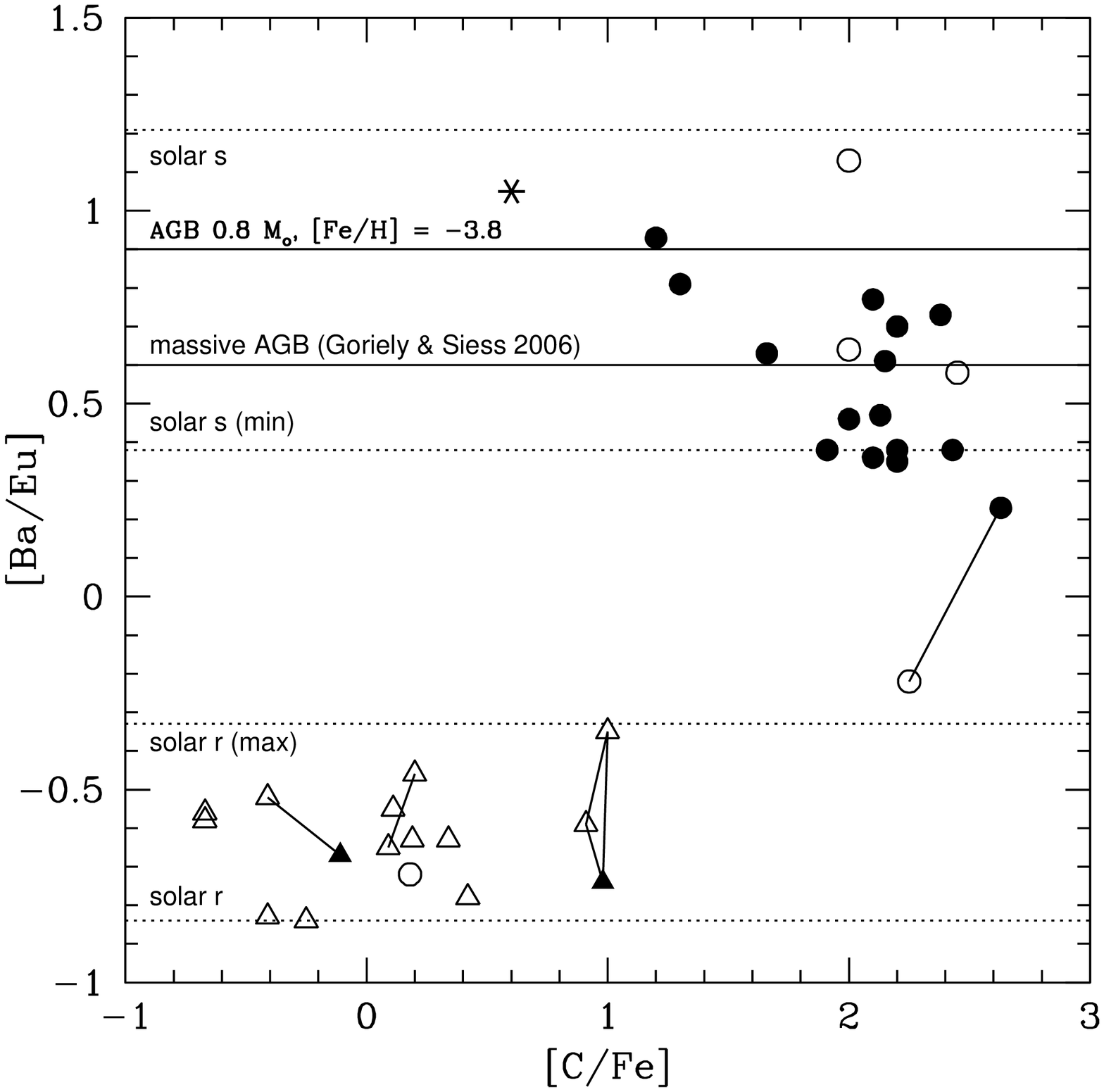}
  \includegraphics[width=8cm]{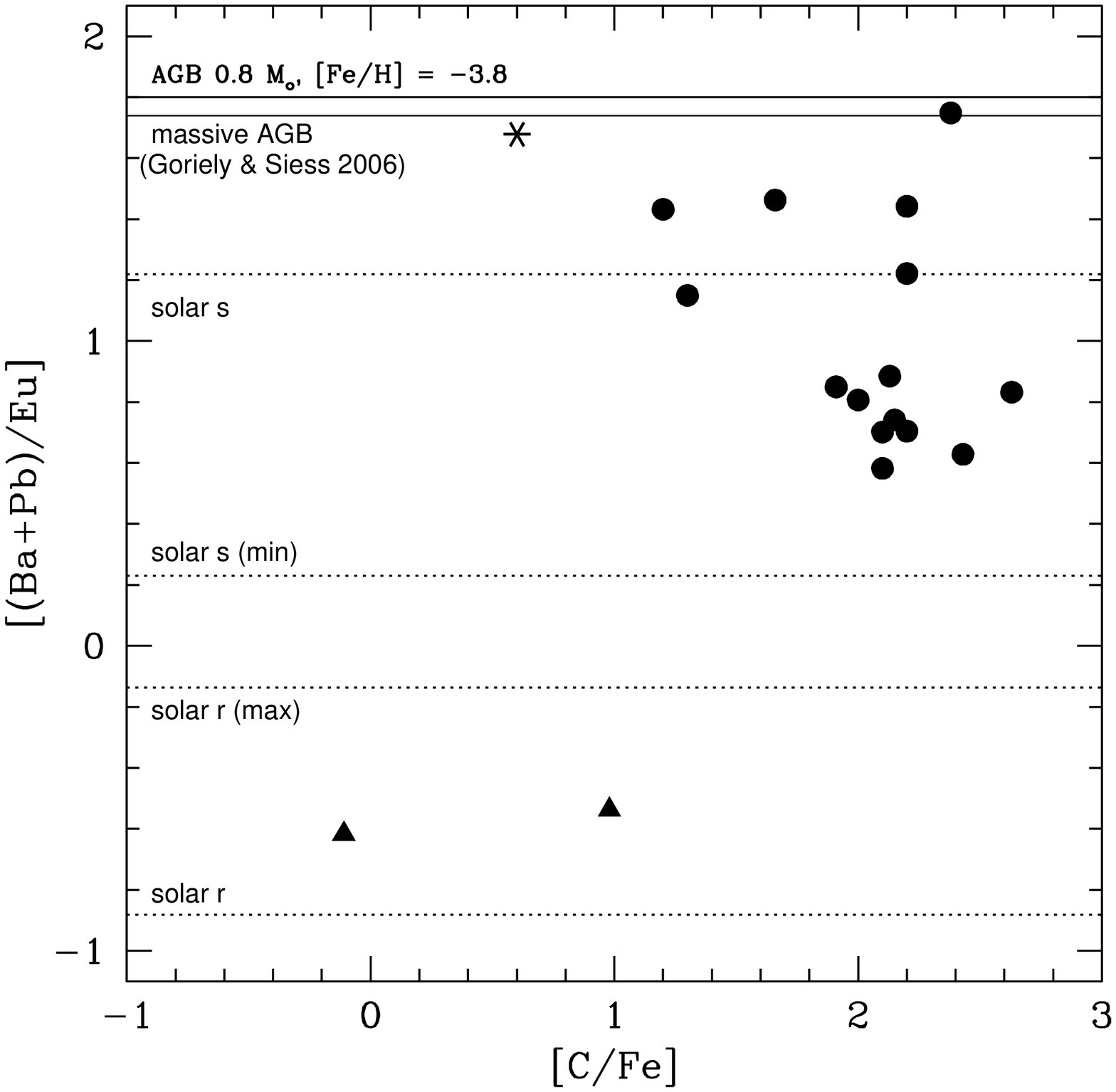}
      \caption{\label{Fig:EuBa}
Upper panel: [Ba/Eu] as a function of [C/Fe]. Lower
	panel: [(Ba+Pb)/Eu] as a function of [C/Fe]. 
Triangles: r-process-enhanced stars; circles: s-process-enhanced stars; 
asterisk: \protect\CS. Open symbols correspond to stars with either no
	Pb data or with only an upper limit
	available.  
The solar [Ba/Eu] and [Pb/Eu] ratios ascribed to 
the r- and s-processes
\citep[from ][]{Goriely-99a} are  displayed as dotted lines, along with the 
maximum r-process and minimum s-process solar ratios, taken from Table~3 of \citet{Goriely-99a}.
Predictions from the 0.8~\Msun, [Fe/H] $=-3.8$ model (see
Fig.~\protect\ref{fig:abund_pattern}) and from the s-process operating
in massive AGB stars \citep[from Fig.~6 of][]{Goriely-Siess-2006} are also
shown. 
%The latter falls in the range of the s+r stars.  
              }
   \end{figure}

\section{Observed abundances and possible models}
\label{Sect:discussion}

In solar-metallicity AGB stars, C is brought to the surface by the third
dredge-up, hence the C/O ratio for a given star is expected to increase steadily with
time. Nitrogen is not enhanced, unless the operation of hot-bottom burning (HBB)
in intermediate-mass AGB stars has converted C into N. The situation in
low-metallicity AGB stars is not as simple, as several specific mixing processes
with potentially large impact on the surface abundances have been found to operate in these
stars. Since CS~30322-023 is believed to be a low-mass, extremely
low-metallicity TP-AGB star, as we have argued in Sect.~\ref{Sect:proba}, we now
briefly describe the various mixing processes in low-metallicity stars that have
been specifically discussed in the literature:   

\begin{itemize}
\item Some deep mixing is believed to occur in low-mass, low-metallicity
  stars on the first giant branch, after the luminosity bump. This
  deep mixing brings to  the surface hydrogen-burning ashes, including
  Na and Al in some instances, and is a signature of hot hydrogen
  burning 
\citep{Charbonnel-1994,Charbonnel-1995,Spite-2006}
\item He-flash-driven deep mixing (He-FDDM) occurs at the tip of the
  RGB when the helium-flash convective zone extends into the tail of
  the H-burning shell
  \citep{Fujimoto-90,Hollowell-90,Fujimoto-00,Schlattl-2001,Schlattl-2002,Iwamoto-2004,Suda-2004}.
  The injection of protons in the region of He-burning produces a H-flash
  and causes the convective region to split. The lower
  convective shell is powered by He burning while the upper one is driven
  by H burning. The structural readjustment subsequent to the H-flash leads
  to a CN-strong star, through the merging of the H-driven convective
  shell (formerly polluted with 3$\alpha$--$^{12}$C) 
  with the outer envelope. It is important to note, however,
  that the atmosphere of such a CN-strong star is not expected to be enriched 
  in s-process elements. If the
  injection of protons in the He flash leads to the production of
  $^{13}$C, and subsequently of s-elements, these heavy elements will remain
  trapped in the He-burning convective region, which the envelope
  never penetrates; 
\item Very similar to the previous scenario is the mixing of protons into the
  thermal pulse at the beginning of the TP-AGB phase, which also  
  leads to a CN-strong star
  \citep{Fujimoto-00,Suda-2004}. In contrast with the previous
  situation, however, this process might produce s-process elements, but this needs to be confirmed (Aikawa, priv. comm.);
%\item a variant of the previous case has been encountered in our
%0.8~\Msun\ model, and does not seem to have been described in previous 
%studies:
%after the first thermal pulse, the 3DUP  engulfs the whole He-burning shell,
%bringing to the surface  essentially $^{12}$C, and some $^{13}$C and $^{14}$N;
\item ``Carbon injection'' is a mixing process encountered
  exclusively in  $Z = 0$ models 
  \citep{Chieffi-2001,Siess-2002}. It occurs at the end of the early AGB phase
  when the He-burning shell (He-BS), while approaching the hydrogen-burning
  shell, is subject to small- amplitude instabilities. In this process,
  after the development of a weak
  thermal pulse in the He-BS, a secondary convective zone forms at the He-H
  discontinuity. It grows in mass, penetrates downwards into the
  C-rich layers left over from
  the pulse, and initiates a H-flash by activation of CNO
  reactions. The subsequent structural readjustment causes 
  the envelope to penetrate in this secondary convective zone, leading
  to a surface CN-enrichment;
\item The ``classical'' proton-mixing scenario assumes that some
  extra mixing takes place at the base of the convective envelope
  during the third dredge-up. It is responsible for the injection of
  protons into the C-rich layers left by the pulse, leading 
  to the formation of a
  $^{13}$C pocket. The radiative burning of $^{13}$C during the
  interpulse phase
  then leads to the production of s-process elements which, after
  passing through the pulse, are brought to the surface by the third dredge-up
  \citep{Straniero-95,Goriely-00}.
\end{itemize} 

In addition to these extra-mixing mechanisms, HBB can also significantly modify
the observed surface abundances, at least in intermediate-mass stars, when H-burning
occurs at the base of the convective envelope, and partly converts $^{12}$C into
$^{14}$N. HBB can also activate the NeNa and MgAl chains, and therefore could be
responsible for $^{23}$Na and $^{27}$Al enrichment. In Pop~III stars, HBB can be
activated in stars with masses as low as $M \approx 2 M_\odot$ \citep{Siess-2002}.

Signatures of these mixing and HBB processes generally are a combined
enhancement of hydrogen-burning products (low $^{12}$C/$^{13}$C, high N, and
sometimes high Na) {\it and} helium-burning products ($^{12}$C and possibly
s-process elements) in the stellar envelope. Since the stellar envelope is
enriched by primary $^{12}$C, C-enhancement factors of several dex are expected.
For example, in the 0.8~\Msun\ model described above, mixing
between the He-flash and the stellar envelope increases the envelope carbon content to
[C/H]$\simeq +1.0$. Such large overabundances are not confirmed by observations
thus far (see Fig.~\ref{Fig:CvsFe}). The abundances in \CS\ are instead
suggestive of CN-cycling of formerly C-rich matter, perhaps through HBB. Since
HBB is not expected to occur in a 0.8~\Msun\ star, we are thus left with
the following alternatives. Either some extra mixing occurs in low-mass stars
that results in  H-burning similar to HBB, or an enrichment by an
intermediate-mass AGB companion which has undergone HBB has polluted the
presently-observed star. The latter suggestion must await firm evidence for
orbital motion, which is lacking at present.

Further constraints on the the nucleosynthetic history of this star come from
the s-process enrichment observed in \CS. The s-process in AGB stars is so far
understood by the partial mixing of protons in C-rich layers, followed by the
3DUP of this material \citep{Straniero-95,Goriely-00}. As shown in
Fig.~\ref{fig:abund_pattern}, the \CS\ s-process abundance pattern is in
excellent agreement with the so-called partial-mixing model. However, in such a
scenario, the s-process surface enrichment is inevitably accompanied by a large
C enrichment due to the 3DUP of primary C made in the thermal pulse, in
contradiction to the low C-to-s-process overabundance exhibited by \CS. None of
the present s-process models could possibly explain such a ratio; for this
reason \CS\ is quite similar to the C-poor s-rich post-AGB star
\object{V453~Oph} observed by  
\citet{Deroo-2005}.

More generally, the observed relation between the [Pb/Ba] ratio and metallicity shown in
Fig.~\ref{Fig:PbBaFeH} is not reproduced by any current models.
The traditional proton-mixing  scenario \citep{Goriely-00} cannot explain the
low [Pb/Ba] stars observed at low metallicity, whereas the mixing of
protons in the thermal pulse at the beginning of the TP-AGB phase,
according to the models of \citet{Fujimoto-00} and \citet{Suda-2004}, predicts a
more complicated trend between [Fe/H] and [Pb/Ba], not seen in Fig.~\ref{Fig:PbBaFeH} 
\citep[compare with Fig.~7 of ][]{Suda-2004}. 
%Similarly, 
%the separation of carbon-rich metal-poor stars in cases II (absence of
%3DUP along the TP-AGB) and
%II' (3DUP along the TP-AGB) 
%as suggested by \citet{Fujimoto-00} and \citet{Suda-2004} is not supported 
%by Fig.~\ref{Fig:BaFevsC+N} \citep[compare with Fig.~6 of ][]{Suda-2004}.

%Therefore, we suggest that the upward diffusion of processed matter accompanying the downward diffusion of protons in the framework of the classical proton-mixing scenario \citep{Goriely-00,Goriely-Siess-01,Busso-01} is able to produce the expected abundance pattern, namely strong overabundances of $^{13}$C, N and s-elements without strong C overabundance.

Our new observation, as well as the C-poor, s-rich post-AGB star of
\cite{Deroo-2005}, clearly show that our understanding of the mixing
mechanism(s) that are responsible, not only for the surface enrichment, but also
for more specific nucleosynthesis events such as the s-process, are still far
from being under control. Our ability to quantitatively explain the existing
observations remains highly dependent on stellar modeling, and most of all on
the description of convective boundaries where some extra mixing takes place. A
better description of the internal transport processes must also be accompanied
with better knowledge of the mass loss, which indirectly influences the
thermodynamical conditions at the base of the convective envelope and
consequently the 3DUP events. 

\section{Summary}

The star \CS\ appears to be the most metal-poor ([Fe/H] = $-3.5$)
s-process-rich star known so far.  It is also a lead star, with [Pb/Ba]
= $+0.95$. An LTE analysis of the Fe~I/Fe~II ionization balance
yields a very low gravity of $\log g = -0.3$. Although it is likely
that such an LTE analysis underestimates the gravity, a comparison
with other CEMP stars analyzed with the LTE assumption 
reveals that \CS\ is the most evolved example among known extremely metal-poor stars. It is
most likely to be a low-mass ($\sim 0.8$~\Msun) TP-AGB star. Larger masses would
put \CS\ too far ($> 20$~kpc) in the outskirsts of the halo, where no
recent star formation is expected. Despite the rather short lifetime
of the TP-AGB phase, the high luminosity of TP-AGB stars makes it
possible to sample them in a very large volume. The probability to
uncover such low-mass, low-metallicity TP-AGB stars is therefore not
small, as confirmed by a detailed quantitative analysis. We thus expect that
additional examples will be revealed in the near future, as spectroscopic surveys of the
Galaxy push to ever fainter limiting magnitudes.

The observed abundance pattern of \CS\ is, however, quite puzzling, in that it exhibits
s-process elements without a strong C overabundance, as would be
expected if the s-process operates in a He-burning environment. The
light-element abundance pattern is instead reminiscent of the
operation of the CN cycle (strong N overabundance, low
$^{12}$C/$^{13}$C ratio, Na overabundance). Hot-bottom burning is
not expected to operate in such low-mass stars. We therefore
suggest that the enrichment of the envelope with nuclear-burning
products does not result from the traditional third dredge-up events,
but rather may be caused by some mixing mechanism that has yet to be identified.
 Although there is no firm evidence at present for orbital
  motion of this star, 
an alternative possibility is that \CS\ resides in a binary
  system, and has been polluted by matter from an intermediate-mass
  companion that produced s-process elements and converted C into N
  during HBB episodes. Clearly, full understanding of the 
abundance pattern and apparent evolutionary state of \CS\ will depend on future 
high-resolution studies of stars with similar characteristics. Abundance determinations duly 
accounting for  NLTE and 3D effects, which are suspected to play an important role in metal-poor giant stars,
should as well bring progress in this field.

\begin{acknowledgements}
We thank J. Johnson, S. Lucatello and M. Spite  
for providing us with supplementary radial-velocity 
measurements, N. Christlieb for a critical reading of the manuscript, and C. 
Laporte for help with a preliminary abundance analysis of \CS.  
D. Pourbaix provided 
help with the orbit search.  T.C.B. acknowledges 
partial funding for this work from grants AST 00-98508, AST 00-98549, and AST 
04-06784, as well as from grant PHY 02-16783: Physics Frontiers Center/Joint
Institute for Nuclear Astrophysics (JINA), all from the U.S. National Science 
Foundation. The {\it Fonds National de la Recherche Scientifique} 
(F.N.R.S., Belgium) supported the research presented in this paper, since 
S.vE., S.G. and L.S. are F.N.R.S. Research Associates,
A.J. is Senior Research Associate and B.F. is Scientific 
Research Worker. 
\end{acknowledgements}

\def\aj{AJ}                   % Astronomical Journal
\def\araa{ARA\&A}             % Annual Review of Astron and Astrophys
\def\apj{ApJ}                 % Astrophysical Journal
\def\apjl{ApJ}                % Astrophysical Journal, Letters
\def\apjs{ApJS}               % Astrophysical Journal, Supplement
\def\ao{Appl.~Opt.}           % Applied Optics
\def\apss{Ap\&SS}             % Astrophysics and Space Science
\def\aap{A\&A}                % Astronomy and Astrophysics
\def\aapr{A\&A~Rev.}          % Astronomy and Astrophysics Reviews
\def\aaps{A\&AS}              % Astronomy and Astrophysics, Supplement
\def\azh{AZh}                 % Astronomicheskii Zhurnal
\def\baas{BAAS}               % Bulletin of the AAS
\def\jrasc{JRASC}             % Journal of the RAS of Canada
\def\memras{MmRAS}            % Memoirs of the RAS
\def\mnras{MNRAS}             % Monthly Notices of the RAS
\def\pra{Phys.~Rev.~A}        % Physical Review A: General Physics
\def\prb{Phys.~Rev.~B}        % Physical Review B: Solid State
\def\prc{Phys.~Rev.~C}        % Physical Review C
\def\prd{Phys.~Rev.~D}        % Physical Review D
\def\pre{Phys.~Rev.~E}        % Physical Review E
\def\prl{Phys.~Rev.~Lett.}    % Physical Review Letters
\def\pasp{PASP}               % Publications of the ASP
\def\pasj{PASJ}               % Publications of the ASJ
\def\qjras{QJRAS}             % Quarterly Journal of the RAS
\def\sci{Science}             % Science
\def\skytel{S\&T}             % Sky and Telescope
\def\solphys{Sol.~Phys.}      % Solar Physics
\def\sovast{Soviet~Ast.}      % Soviet Astronomy
\def\ssr{Space~Sci.~Rev.}     % Space Science Reviews
\def\zap{ZAp}                 % Zeitschrift fuer Astrophysik
\def\nat{Nature}              % Nature
\def\iaucirc{IAU~Circ.}       % IAU Cirulars
\def\aplett{Astrophys.~Lett.} % Astrophysics Letters
\def\apspr{Astrophys.~Space~Phys.~Res.}
                % Astrophysics Space Physics Research
\def\bain{Bull.~Astron.~Inst.~Netherlands} 
                % Bulletin Astronomical Institute of the Netherlands
\def\fcp{Fund.~Cosmic~Phys.}  % Fundamental Cosmic Physics
\def\gca{Geochim.~Cosmochim.~Acta}   % Geochimica Cosmochimica Acta
\def\grl{Geophys.~Res.~Lett.} % Geophysics Research Letters
\def\jcp{J.~Chem.~Phys.}      % Journal of Chemical Physics
\def\jgr{J.~Geophys.~Res.}    % Journal of Geophysics Research
\def\jqsrt{J.~Quant.~Spec.~Radiat.~Transf.}
                % Journal of Quantitiative Spectroscopy and Radiative Trasfer
\def\memsai{Mem.~Soc.~Astron.~Italiana}
                % Mem. Societa Astronomica Italiana
\def\nphysa{Nucl.~Phys.~A}   % Nuclear Physics A
\def\physrep{Phys.~Rep.}   % Physics Reports
\def\physscr{Phys.~Scr}   % Physica Scripta
\def\planss{Planet.~Space~Sci.}   % Planetary Space Science
\def\procspie{Proc.~SPIE}   % Proceedings of the SPIE

\let\astap=\aap
\let\apjlett=\apjl
\let\apjsupp=\apjs
\let\applopt=\ao

%\bibliographystyle{aa}

%\begin{thebibliography}{}

%\end{thebibliography}

%\bibliography{ajorisse_articles}

\end{document}